\documentclass[a4paper,notitlepage,eqsecnum,nofootinbib]{revtex4-1}
\usepackage{slashed,bbold,bm,amsmath,amssymb,mathtools,wrapfig,epsfig} 
\usepackage[usenames]{color}
\usepackage{hyperref}
\usepackage{tcolorbox}
\hypersetup{
    colorlinks=true,       % false: boxed links; true: colored links
    linkcolor=red,          % color of internal links (change box color with linkbordercolor)
    citecolor=blue,        % color of links to bibliography
    filecolor=green,      % color of file links
    urlcolor=magenta           % color of external links
}

\newcommand{\crt}{\\[2mm]}
\newcommand{\nn}{\nonumber}
\newcommand{\beq} {\begin{equation}}
\newcommand{\eeq} {\end{equation}}
\newcommand{\beqa} {\begin{eqnarray}}
\newcommand{\eeqa} {\end{eqnarray}}

\newcommand{\bs}[1]{\boldsymbol{#1}}

\newcommand{\ie}{{\it i.e.}}

\newcommand{\eg}{{\it e.g.}}

\newcommand{\as}{\alpha_s}
\newcommand{\lqcd}{\Lambda_{QCD}}
\newcommand{\la}{\Lambda}
\newcommand{\lm}{\lambda}

\newcommand{\vphi}{\varphi}
\newcommand{\veps}{\varepsilon}
\newcommand{\order}[1]{${\cal O}\left(#1 \right)$}
\newcommand{\morder}[1]{{\cal O}\left(#1 \right)}
\newcommand{\eq}[1]{(\ref{#1})}
\newcommand{\fig}[1]{Fig.~\ref{#1}}
\newcommand{\inv}[1]{\frac{1}{#1}}
\newcommand{\halft}{{\textstyle \frac{1}{2}}}
\newcommand{\quart}{{\textstyle \frac{1}{4}}}

\newcommand{\intt}{{\textstyle \int}}
\newcommand{\ket}[1]{\left\vert{#1}\right\rangle}
\newcommand{\ketb}[1]{\vert{#1}\rangle}
\newcommand{\bra}[1]{\langle{#1}\vert}

\newcommand{\com}[2]{\left[{#1},{#2}\right]}
\newcommand{\comb}[2]{\big[{#1},{#2}\big]}
\newcommand{\acom}[2]{\left\{{#1},{#2}\right\}}
\newcommand{\acomb}[2]{\big\{{#1},{#2}\big\}}
\newcommand{\tr}{\mathrm{Tr}\,}

\newcommand{\mE}{\mathcal{E}}

\newcommand{\mH}{\mathcal{H}}

\newcommand{\mJ}{\mathcal{J}}
\newcommand{\mK}{\mathcal{K}}

\newcommand{\xv}{{\bs{x}}}

\newcommand{\yv}{{\bs{y}}}
\newcommand{\zv}{{\bs{z}}}

\newcommand{\Av}{{\bs{A}}}

\newcommand{\Ev}{{\bs{E}}}
\newcommand{\Jv}{{\bs{J}}}

\newcommand{\Lv}{{\bs{L}}}
\newcommand{\Pv}{{\bs{P}}}
\newcommand{\Pva}{{\bs{P}_A}}
\newcommand{\Pvb}{{\bs{P}_B}}

\newcommand{\Sv}{{\bs{S}}}

\newcommand{\gz}{\gamma^0}

\newcommand{\gf}{\gamma_5}

\newcommand{\Ga}{\Gamma}

\newcommand{\rar}{\rightarrow}
\newcommand{\lar}{\leftarrow}

\newcommand{\rder}{{\buildrel\rar\over{\partial}}}
\newcommand{\lder}{{\buildrel\lar\over{\partial}}}

\newcommand{\nv}{\bs{\nabla}}

\newcommand{\rnab}{{\overset{\rar}{\nv}}\strut}
\newcommand{\lnab}{{\overset{\lar}{\nv}}\strut}
\newcommand{\rhap}{{\overset{\rar}{H}_0}\strut^{\hspace{-.5em}\scriptscriptstyle{(\Pv)}}}
\newcommand{\lhap}{{\overset{\lar}{H}_0}\strut^{\hspace{-.5em}\scriptscriptstyle{(\Pv)}}}

\newcommand{\xtr}{\bs{x}_\perp}
\newcommand{\xt}{x_\perp}

\newcommand{\alv}{{\bs{\alpha}}}

\newcommand{\aly}{{\alpha_2}}
\newcommand{\alz}{{\alpha_3}}
\newcommand{\at}{\alpha_\perp}
\renewcommand{\ap}{\alpha_\vphi}  % for revtex

\newcommand{\phip}{\phi^{\scriptscriptstyle{(P)}}}
\newcommand{\phir}{\phi^{\scriptscriptstyle{(0)}}}

\newcommand{\Phir}{\Phi^{\scriptstyle{(0)}}}
\newcommand{\Phip}{\Phi^{\scriptscriptstyle{(\Pv)}}}
\newcommand{\Phiz}{\Phi^{\scriptscriptstyle{(P)}}}

\newcommand{\dr}{\partial_r}
\newcommand{\dz}{\partial_z}
\newcommand{\dpp}{\partial_\perp}

\newcommand{\dt}{\partial_t}
\newcommand{\dxi}{\delta\xi}
\newcommand{\dphi}{\partial_\vphi}

\begin{document}
%{\par\raggedleft \texttt{\vspace{-.3cm} Boost_23.tex}\par}

\title{QCD bound states in motion}

\author{Paul Hoyer}
\affiliation{ \vspace{1mm} Department of Physics, POB 64, FIN-00014 University of Helsinki, Finland}
\email{paul.hoyer@helsinki.fi}

\begin{abstract} 
I consider the frame dependence of QCD bound states in the presence of a confining, spatially constant gluon field energy density. The states are quantized at equal time in $A^0=0$ (temporal) gauge. I derive the frame dependence of the wave functions, and demonstrate the Lorentz covariance of the electromagnetic (transition) form factors for states of any spin. The wave functions of $J^{PC}=0^{-+}$ states with CM momentum $\Pv \neq 0$ are considered in some detail, verifying their local normalizability and the expected frame dependence of the bound state energy. 
\end{abstract}

\maketitle

\parindent 0cm
%\vspace{-.2cm}

%%%%%%%%%%%%%%%%%%%%%%%%%%
\section{Remarks on bound states} \label{sec1}
%%%%%%%%%%%%%%%%%%%%%%%%%%

Atoms and hadrons are our primary examples of physical bound states. Their field theory (QED and QCD) description complements the methods of scattering amplitudes, but is not part of the standard QFT curriculum. Bound states are challenging, but the omission even of their general principles seems unwarranted.

Analytic evaluations of Standard Model dynamics mainly rely on expansions in powers of the coupling. The lowest order term should provide an adequate first approximation of the full result. Scattering amplitudes reduce to free propagators at vanishing coupling, which defines the lowest order of the perturbative $S$-matrix. 

Bound state constituents interact at all times. Bound states are typically expanded around solutions of the Schr\"odinger or Bethe-Salpeter type equations. The choice of initial (non-perturbative) state affects its higher order perturbative corrections such that the full series (for physical quantities) is formally independent of the initial choice \cite{Caswell:1978mt,Lepage:1978hz}. 

Each order of a perturbative expansion must have Poincar\'e symmetry. The explicit covariance of Feynman diagrams is enabled by their free propagators. Bound states are eigenstates of the Hamiltonian, which is frame dependent. Consequently  bound state covariance is not fully explicit (kinetic), but also realized dynamically (through interactions). \textit{E.g.,} time translation invariance is ensured for eigenstates of the Hamiltonian, which includes interactions.

In equal-time quantization the Hamiltonian commutes with space translations and rotations, but not with boosts. Bound state masses and quantum numbers can be determined in the rest frame \cite{RevModPhys.57.723,Caswell:1985ui,Adkins:2015wya,Adkins:2018lvj}, whereas bound state scattering involves moving states. The frame dependence of atoms is dynamic and non-trivial \cite{Brodsky:1968xc,Brodsky:1968ea,PhysRev.176.1523}.

Hamiltonians quantized at equal light-front (LF) time $x^+ \equiv t+z$ commute with boosts \cite{Burkardt:1995ct,Brodsky:1997de}. Photons propagating in the negative $z$-direction interact at equal $x^+$, making LF wave functions advantageous for describing form factors. Rotation symmetry is realized dynamically on the LF since $x^+$ depends on $z$. This complicates the determination of angular momentum (except $J^z$), and raises some delicate issues \cite{Mannheim:2020rod,Polyzou:2023vjj}. 

In the following I consider equal-time quantization, which is well established and allows to determine the $J^{PC}$ quantum numbers in the rest frame. The dynamic realization of boost covariance is the main topic of this paper, and serves as a non-trivial check of the method.

The equal-time wave function of Positronium in motion was determined in \cite{Jarvinen:2004pi} (see also \cite{Hoyer:2021adf}). The $\ket{e^+e^-}$ Fock state Lorentz contracts as expected. The increase of the Coulomb potential due to the contraction is cancelled by the $\ket{e^+e^-\gamma}$ Fock contribution, which (at leading order) vanishes in the rest frame.

The physical hadron scale $\lqcd \simeq$ 1 fm$^{-1}$ is not a parameter of the QCD Lagrangian. The scale appears in renormalization, and determines long-distance features such as hadron radii. Color confinement is not apparent even in formally exact methods such as Dyson-Schwinger equations \cite{Ding:2022ows} and NRQCD \cite{Brambilla:2004jw}. Expansions based on initially unconfined quark and gluon states may not recover confinement.

I introduce the QCD scale through a \textit{boundary condition}, which preserves the equations of motion. Temporal $(A_a^0 = 0)$ gauge fixing imposes Gauss' law on the classical longitudinal electric field. For (globally) color singlet states a specific homogenous solution of Gauss' law can be included \cite{Hoyer:2021adf}. This gives rise to a spatially constant gluon field energy density, characterized by a universal scale parameter $\la$. Setting $\la=0$ gives the standard Coulomb potential, whereas $\la>0$ adds an instantaneous potential, which is linear for $q\bar q$ states. 

The confining potential provides a rich dynamics for the color singlet bound states even at $\as=0$. This may serve as the lowest order of a perturbative expansion in $\as$ provided it respects the exact symmetries of QCD, including Poincar\'e invariance. Here I verify that the bound states and form factors have the correct frame dependence. These properties are dynamically realized in equal-time quantization, and have not been demonstrated before.

In section \ref{sec2} I recall the bound state method and some results presented in \cite{Hoyer:2021adf}. Section \ref{sec3} considers how the bound states transform under infinitesimal boosts, and verifies the Lorentz covariance of their form factors. Section \ref{sec4} presents a general Dirac basis for the wave functions, which allows to express the bound state equation as a set of coupled partial differential equations. As a specific example, section \ref{sec5} considers properties of the $J^{PC}=0^{-+}$ wave functions, using both analytic and numerical methods. Summary remarks are given in section \ref{sec6}.

%%%%%%%%%%%%%%%%%%%%%%%%%%
\section{The bound state method \cite{Hoyer:2021adf}} \label{sec2}
%%%%%%%%%%%%%%%%%%%%%%%%%%

%%%%%%%%%%%%%%%%%%%%%%%%%%
\subsection{Temporal gauge} \label{sec2B}
%%%%%%%%%%%%%%%%%%%%%%%%%%

Gauge theory actions have no $\dt A^0$ nor $\nv\cdot\Av$ terms. Hence the $A^0$ and longitudinal $\Av_L$ fields do not propagate in space-time. For scattering amplitudes it is convenient to add the missing terms through a covariant gauge fixing Lagrangian $\propto(\partial_\mu A^\mu)^2$, achieving explicit Poincar\'e invariance. For bound states the instantaneous gauge interactions due to $A^0$ and $\Av_L$ actually are welcome.

Bound state calculations commonly use Coulomb gauge ($\nv\cdot\Av=0$). The absence of a field conjugate to $A^0$ causes well-known complications \cite{Christ:1980ku,Weinberg:1995mt}. Here I choose temporal gauge ($A^0=0$) \cite{Willemsen:1977fr,Bjorken:1979hv,Leibbrandt:1987qv,Strocchi:2013awa}, in which canonical quantization is straightforward. To my knowledge there are no previous bound state calculations in temporal gauge.

The condition $A_a^0=0$ does not exclude time independent gauge transformations, which are generated by Gauss operator $G_a(t,\xv) \equiv \delta S_{QCD}/\delta A^0_a(t,\xv)$. In temporal gauge $G_a=0$ is not an operator equation of motion, but is imposed as a \textit{constraint on physical states}: $G_a(t,\xv)\ket{phys}=0$. This ensures the invariance of the physical states under the remaining gauge degrees of freedom. The constraint is invariant under time evolution since $G_a$ commutes with the Hamiltonian. It requires the longitudinal electric field $\Ev_L = -\dt\Av_L$ of a physical state to satisfy
\begin{align} \label{eII29}
\nv\cdot \Ev_{L}^{a}\ket{phys} = g\big[- f_{abc}\Av_b\cdot \Ev_c+\psi^\dag T^a\psi\big]\ket{phys} \equiv g\,\mE_a\ket{phys}
\end{align}
where $f_{abc}$ are the structure constants and $T^a$ the generators in the fundamental representation of the color SU(3) group.

The standard boundary condition in solving \eq{eII29} is $\Ev_{L}^{a}(t,\xv\to\infty)=0$, giving the instantaneous Coulomb potential. For (globally) color singlet states in QCD I include a homogeneous ($\nv\cdot\Ev_{L}^{a}(t,\xv)=0$) solution,
\begin{align} \label{eII31}
\Ev_{L}^a(t,\xv)\ket{phys} &= -\nv_x \int d\yv \Big[\kappa\,\xv\cdot\yv + \frac{g}{4\pi|\xv-\yv|}\Big]\mE_a(t,\yv) \ket{phys}
\end{align}
where $\mE_a$ is defined in \eq{eII29} and the normalization $\kappa$ of the homogeneous solution is independent of $\xv$ and $\yv$. $\Ev_{L}^{a}$ contributes an instantaneous interaction $\mH_V(t)$ to the QCD Hamiltonian acting on $\ket{phys}$,
\begin{align} \label{eII32}
\mH_V(t) \equiv \halft\int d\xv\,\big(\Ev_{L}^a\big)^2 = \int d\yv d\zv\Big[\,\yv\cdot\zv \big(\halft\kappa^2\intt d\xv + g\kappa\big) + \halft \frac{\as}{|\yv-\zv|}\Big]\mE_a(t,\yv)\mE_a(t,\zv)
\end{align} 

%%%%%%%%%%%%%%%%%%%%%%%%%%
\subsection{The instantaneous potential} \label{sec2C}
%%%%%%%%%%%%%%%%%%%%%%%%%%

A globally color singlet $q\bar q$ state of rest mass $M$ and momentum $\Pv$ can at time $t$ be expressed as
\begin{align}  \label{qcd97}
\ket{M,\Pv} = \inv{\sqrt{N_c}}\sum_{A,B}\sum_{\alpha,\beta}\int d\xv_1 d\xv_2\,\bar\psi_\alpha^A(t,\xv_1)e^{i\Pv\cdot(\xv_1+\xv_2)/2}\delta^{AB}\Phi^{(\Pv)}_{\alpha\beta}(\xv_1-\xv_2)\psi_\beta^B(t,\xv_2)\ket{0}
\end{align}
where $N_C=3$ for QCD, $A,B$ are color and $\alpha,\beta$ Dirac indices. The $\bar\psi$ field creates a quark at $(t,\xv_1)$ and $\psi$ simul\-taneously an antiquark at $(t,\xv_2)$. 
The plane wave phase $\exp[i\Pv\cdot(\xv_1+\xv_2)/2]$ ensures the correct translation dependence of the entire bound state, while the $c$-numbered wave function $\Phi^{(\Pv)}_{\alpha\beta}(\xv_1-\xv_2)$ determines the relative distribution of the (single flavored) quarks. Each component $\ket{q(\xv_1)\bar q(\xv_2)} \equiv \sum_A\bar\psi^A_\alpha(t,\xv_1)\psi^A_\beta(t,\xv_2)\ket{0}$ is a physical state in temporal gauge,
whose instantaneous field $\Ev_L$ is determined by \eq{eII31}. 
The state is gauge dependent, and defines observables such as four-momenta, form factors and scattering amplitudes which must be gauge-invariant.

The term $\propto \kappa^2\intt d\xv$ in \eq{eII32} shows that the homogeneous solution brings an $\xv$-independent field energy density. The total energy ($\propto$ the volume of space) is irrelevant only if it is universal, \ie, the same for all physical states. The normalization $\kappa$ depends on the state and is determined by the requirement that the field energy density be universal.

For the component $\ket{q(\xv_1)\bar q(\xv_2)}$ the \order{\kappa^2} and \order{g\kappa} terms in $\mH_V(t)$ \eq{eII32} give contributions proportional to,
\begin{align} \label{e2b1a} 
\int d\yv\, d\zv\,\yv\cdot\zv\,\mE_a(t,\yv)\mE_a(t,\zv)\ket{q(\xv_1)\bar q(\xv_2)}=C_F (\xv_1-\xv_2)^2\ket{q(\xv_1)\bar q(\xv_2)}
\end{align}
Universality of the energy density requires $\kappa^2 \propto 1/(\xv_1-\xv_2)^2$ for $\ket{q(\xv_1)\bar q(\xv_2)}$. The \order{g\kappa} contribution is then $\propto |\xv_1-\xv_2|$. Denoting the field energy density as $E_\la = \la^4/(2g^2C_F)$ and omitting its universal (infinite) contribution,
\begin{align} \label{e2b1}
\mH_V(t)\ket{q(\xv_1)\bar q(\xv_2)} =\Big(\la^2\,|\xv_1-\xv_2| - C_F\frac{\as}{|\xv_1-\xv_2|}\Big)\ket{q(\xv_1)\bar q(\xv_2)} \equiv V_{q\bar q}(|\xv_1-\xv_2|)\ket{q(\xv_1)\bar q(\xv_2)} 
\end{align}
where $C_F=(N_c^2-1)/2N_c =4/3$ for QCD. $V_{q\bar q}(|\xv_1-\xv_2|)$ agrees with the phenomenologically determined ``Cornell potential'' for quarkonia \cite{Eichten:1979ms,Eichten:2007qx}. It is applicable also to relativistic light quark states, being independent of the quark masses and momenta.
Confining potentials analogous to \eq{e2b1} are found for any color singlet quark and gluon state. 

The full QCD Hamiltonian acting on the $\ket{q\bar q}$ state \eq{qcd97} can create a transversely polarized gluon. Similarly  $\mH_{QCD}\ket{q\bar qg}$ contains $\ket{q\bar qgg}$ and $\ket{q\bar q\,q\bar q}$. Higher Fock states are suppressed by powers of the coupling $g$. This defines a formally exact ``bound Fock expansion'' for mesons,
\begin{align} \label{e2b2} 
\ket{\textit{Meson}} = C_{q\bar q}\ket{q\bar q}_I+ C_{q\bar qg}\ket{q\bar qg}_I+ \ldots + C_{q\bar q\,q\bar q}\ket{q\bar q\,q\bar q}_I+ \ldots
\end{align}
where $C_{q\bar q}$ is \order{g^0},  $C_{q\bar q g}$ is \order{g}, $C_{q\bar q\,q\bar q}$ is \order{g^2}, ... 
The subscript $I$ indicates that the Fock constituents interact through their instantaneous potential, \eg, \eq{e2b1} for $\ket{q(\xv_1)\bar q(\xv_2)}$.
The number of Fock states is limited by the orders of $g$ included in the perturbative expansion. Annihilations between the constituents, \eg, $\ket{q\bar qg}_I \to \ket{q\bar q}_I$ contribute higher order corrections to Fock states with fewer constituents.

%%%%%%%%%%%%%%%%%%%%%%%%%%
\subsection{\order{\as^0} bound states at rest} \label{sec2D}
%%%%%%%%%%%%%%%%%%%%%%%%%%

In a perturbative approach already the lowest \order{\as^0} bound states should provide a reasonable approximation of physical hadrons. With the valence $\ket{q\bar q}_I$ \eq{qcd97} chosen as initial state the coefficients of all other bound Fock components in \eq{e2b2} vanish at $\as=0$, and the potential in \eq{e2b1} is linear,
\begin{align} \label{2b4} 
V_{q\bar q}(|\xv_1-\xv_2|)\Big|_{\as=0} \equiv V(r) = \la^2 r \equiv V'r
\end{align}
Including the free quark Hamiltonian,
\begin{align} \label{2b4a} 
\mH_{QCD}(t=0) = \mH_0+ \mH_{V} \hspace{2cm}
\mH_0 = \int d\xv\,\psi^\dag(0,\xv)\,H_0\,\psi(0,\xv)\ \hspace{.8cm} H_0 = -i\alv\cdot\rnab+m\gz
\end{align}
the eigenstate condition $\mH_{QCD}\ket{M,\Pv=0}= M\ket{M,\Pv=0}$ imposes a bound state equation (BSE) on the rest frame wave function $\Phir \equiv \Phi$,
\begin{align} \label{qcd42}
\big(i\alv\cdot\rnab+m\gz\big)\Phi(\xv)+\Phi(\xv)\big(i\alv\cdot\lnab-m\gz\big) = \big[M- V(r)\big]\Phi(\xv)
\end{align}
where $\xv=\xv_1-\xv_2$ and $r=|\xv|$. Rotation symmetry allows to classify the states according to their eigenvalues $j(j+1)$ and $\lambda$ of the angular momentum operators $\bs{\mJ}^2$ and $\mJ^z$, where (suppressing color and $t=0$)
\begin{align} \label{qcd49}
\bs{\mJ} &= \int d\xv\,\psi^\dag(\xv)\,\bs{J}\,\psi(\xv) \hspace{6cm} \bs{J} = \Lv+\Sv= \xv\times(-i\nv)+\halft\gf\alv \nn\crt
\bs{\mJ}\ket{M,\Pv=0}&=\int d\xv_1 d\xv_2\, \bar\psi(\xv_1) \com{\Jv}{\Phi^{j\lm}(\xv_1-\xv_2)}\psi(\xv_2)\ket{0} \hspace{2cm} \Phi_{\alpha\beta}^{j\lm}(\xv) = \sum_{i=1}^{16} \Gamma_{\alpha\beta}^{(i)}F_i(r)Y_{j\lambda}(\hat\xv)
\end{align}
The $Y_{j\lambda}(\xv/r)$ are standard spherical harmonics, $F_i(r)$ are ($j$-dependent) radial functions and the $\Gamma_{\alpha\beta}^{(i)}$ are 16 Dirac structures. A convenient choice for the Dirac structures is $1,\ \alv\cdot\xv,\ \alv\cdot\Lv$ and $\alv\cdot\xv\times\Lv$, each multiplied by $1,\gz,\gf$ or $\gz\gf$. The parity $\eta_P$ and charge conjugation $\eta_C$ of the state restricts the allowed Dirac structures.

In the non-relativistic quark model $\eta_P=(-1)^{L+1}$ and $\eta_C=(-1)^{L+S}$, which excludes states with $\eta_P=-\eta_C = (-1)^j$. These values of $\eta_P$ and $\eta_C$ are compatible only with the structures $\gz$ and $\gf\,\alv\cdot\Lv$, which do not contribute to the BSE \eq{qcd42}. Hence the quark model exotics are absent also in the present relativistic framework.

The wave function of physical states with $\eta_P=-\eta_C = (-1)^{j+1}$ and $J^z=\lm$ can be expressed as,
\begin{align} \label{qcd57}
\Phi^{j\lambda}(\xv) = \Big[F_1(r) + i\,\alv\cdot\xv\,F_2(r) + \alv\cdot\xv\times\Lv\,F_3(r) + \gz\,F_4(r)\Big]\gf\,Y_{j\lambda}(\hat\xv)
\end{align}
Substituting this into the BSE \eq{qcd42} relates the radial functions $F_i(r)$ and imposes a radial equation on $F_1(r)$,
\begin{align} \label{qcd59}
F_2(r) = \frac{2}{r(M-V)}\partial_r F_1(r) \hspace{2cm} F_3(r) &= \frac{2}{r^2(M-V)}F_1(r)
 \hspace{2cm} F_4(r) = \frac{2m}{M-V}F_1(r) \nn\crt
F_1''+\Big(\frac{2}{r}+\frac{V'}{M-V}\Big)F_1' &+ \Big[\quart (M-V)^2-m^2-\frac{j(j+1)}{r^2}\Big]F_1 = 0
\end{align}
The wave function may be expressed in terms of $F_1(r)$ as
\begin{align}\label{qcd60}
\Phi^{j\lambda}(\xv) &= \Big[\frac{2}{M-V}(i\alv\cdot\rnab+m\gz)+1\Big]\gf\,F_1(r)Y_{j\lambda}(\hat\xv)
\end{align}
and its normalizing integral is
\begin{align}\label{qcd113}
\int d\xv\,\tr\Big[\Phi^{j\lambda^{\scriptstyle{\dag}}}(\xv)\Phi^{j\lambda}(\xv)\Big] = 8\int_0^\infty dr\,r^2F_1^*(r)\Big[1-\frac{2V'}{(M-V)^3}\partial_r\Big]F_1(r)
\end{align}

The behavior of $F_1(r\to\infty)$ given by the radial equation \eq{qcd59} is, up to a phase convention and normalization $N$,
\begin{align} \label{qcd114}
F_1(r\to\infty) \simeq N\,r^{-1-i m^2/V'}\,\exp\big[i(M-V)^2/4V'\big]
\end{align} 
which satisfies the radial equation at \order{r^{-2}} (or \order{r^{-3}} for $m=0$).
Hence the integrand in \eq{qcd113} approaches a constant as $r\to \infty$.
This is a feature also of the Dirac equation with a linear potential \cite{Plesset:1930zz}, and is suggestive of pair production (string breaking). A $q\bar q$ state $\ket{A}$ defined as in \eq{qcd97} has a non-vanishing overlap with a pair of $q\bar q$ states $B,C$: $\bra{B\,C}A\rangle \neq0$, much as depicted in dual diagrams \cite{Harari:1981nn,Rosner:1981np,Zweig:2015gpa}. The feature where an earlier stage of a process $\ket{A}$ averages a later stage $\ket{B\,C}$ is observed in $e^+e^- \to hadrons$ and referred to as quark-hadron duality \cite{Melnitchouk:2005zr}.

The normalizability of non-relativistic Schr\"odinger wave functions determines their discrete energy eigenvalues. Here the analogous requirement for relativistic wave functions is that they should be locally normalizable. In \eq{qcd59} $F_1(r\to 0)\propto r^\beta$ with $\beta = j$ or $\beta = -j-1$. The radial equation determines the regular ($\beta = j$) solution for all $r$, up to its overall normalization. As seen from \eq{qcd60} $\Phi^{j\lambda}(\xv)$ is singular at $V(r)=M$ unless $F_1(r=M/V')=0$. Imposing this determines the allowed bound state masses $M$. For small quark masses $m$ the bound states lie on nearly linear Regge trajectories with evenly spaced daughters (see Fig. 21 of \cite{Hoyer:2021adf}).

%%%%%%%%%%%%%%%%%%%%%%%%%%
\section{Frame dependence} \label{sec3}
%%%%%%%%%%%%%%%%%%%%%%%%%%

The frame dependence of physical observables is determined by Poincar\'e symmetry. The symmetry is realized dynamically, as the generators of boosts do not commute with the Hamiltonian. The Poincar\'e Lie algebra ensures that the boosted states of mass $M$ and 3-momentum $\Pv$ are eigenstates of the Hamiltonian with eigenvalue $E=\sqrt{\Pv^2+M^2}$\,, \ie, they satisfy their bound state equation (BSE). This allows to determine the $\Pv$-dependence of the wave function from the BSE, instead of boosting.

Boosting equal-time $\ket{f\bar f}$ states such as \eq{qcd97} causes unequal time shifts of the constituents (see \eq{infboost} below). The constituents may be returned to equal time through time translations generated by the Hamiltonian. It is straightforward to determine the boost generators of free fermions (see \eq{zboost} and \eq{freeh1} below, with $V=0$). In the absence of interactions the $\ket{f\bar f}$ wave function Lorentz contracts as in classical relativity, see Eq. (8.101) of \cite{Hoyer:2021adf}.

The interactions between bound state constituents affect their time translations. Unless the Hamiltonian is determined by a Poincar\'e invariant field theory its energy eigenvalues need not have the required momentum dependence, see \cite{Artru:1983gm} for an example. Poincar\'e covariance should be ensured for the interaction Hamiltonian $\mH_V$ \eq{eII32}, since it is derived from the QCD action in temporal gauge, including a homogeneous solution of Gauss' constraint. The state \eq{qcd97} is an eigenstate of the \order{\as^0} Hamiltonian \eq{2b4a} if its wave function satisfies the BSE
\begin{align} \label{qcd98}
i\nv\cdot\acomb{\alv}{\Phip(\xv)}-\halft \comb{\alv\cdot\Pv}{\Phip(\xv)}+m\comb{\gz}{\Phip(\xv)} &= \big[E-V(|\xv|)\big]\Phip(\xv)
\end{align}
where $V(|\xv|)=V'|\xv|$ \eq{2b4}. The term $\propto \Pv$ breaks full rotational invariance, but preserves rotational symmetry around $\Pv$. Hence the helicity $\lm$ of the state can be taken to be independent of $\Pv$ (and is suppressed in the following). 

The transformation of states under boosts depends on the gauge, and boosts generally must be combined with a gauge transformation to maintain the original gauge (see \cite{Dietrich:2012iy} for an example). Poincar\'e covariance is required for physical observables, \eg, in \eq{qcd98} we must have $E=\sqrt{\Pv^2+M^2}$\,. The electromagnetic (transition) form factor for $\gamma^*\, A \to B$ is also an observable,
\begin{align}  \label{f2}
F_{AB}^\mu(y) &= \bra{M_B, \Pvb}j^\mu(y)\ket{M_A,\Pva} \hspace{2cm} j^\mu(y) = \bar\psi(y)\gamma^\mu\psi(y)
\end{align}
The form factor was shown \cite{Dietrich:2012un} to be gauge invariant for states with wave functions which satisfy \eq{qcd98}. Here I shall demonstrate that $F_{AB}^\mu(y)$ behaves as a four-vector under Lorentz boosts. This is a stringent test of Poincar\'e symmetry for strongly bound states, recalling the discussion of atomic form factors in \cite{Brodsky:1968xc,Brodsky:1968ea,PhysRev.176.1523}.

%%%%%%%%%%%%%%%%%%%%%%%%%%
\subsection{Infinitesimal boost} \label{sec3A}
%%%%%%%%%%%%%%%%%%%%%%%%%%

A fermion field at $t=0$ is transformed by unitary operators under time translations and boosts,
\begin{align} \label{zboost}
{\rm Time\ translation:}\hspace{.5cm} &U_\mH(t)\psi(0,\xv)U_\mH^\dag(t) = \psi(t,\xv) 
\hspace{4.3cm} U_\mH(\delta t)=1+i\delta t\,\mH+\morder{\delta t^2} \nn \crt
\mbox{Boost\ in\ $z$-direction:}\hspace{.5cm} &U_\mK(\xi)\psi(0,\xtr,z)U_\mK^{\dag}(\xi) = e^{-\xi\alz/2}\psi(z\sinh\xi,\xtr,z\cosh\xi)
\end{align}
An infinitesimal boost $U_\mK(\dxi)$ along the $z$-axis of the state $\ket{M,\Pv}$ \eq{qcd97} shifts the fermion fields to unequal times, 
\begin{align} \label{infboost} 
U_\mK(\dxi)\ket{M,\Pv} = 
 \int d\xv_1\,d\xv_2\,\bar\psi(\dxi\, z_1,\xv_1)\,e^{\dxi\alz/2}\,e^{i\Pv\cdot (\xv_1+\xv_2)/2}\,\Phip(\xv_1-\xv_2)\,e^{-\dxi\alz/2}\,\psi(\dxi\, z_2,\xv_2)\ket{0} + \morder{\dxi^2}
\end{align}
The fields can be shifted back to equal times using time translations \eq{zboost} generated by the Hamiltonian. For $\as=0$ with $\mH_{QCD}$ given by \eq{2b4a},
\begin{align} \label{freeh1} 
\bar\psi(\dxi\, z_1,\xv_1) &= U_{\mH}(\dxi\, z_1)\bar\psi(0,\xv_1)U_{\mH}^\dag(\dxi\, z_1)
= \bar\psi(0,\xv_1)\big[1+(-i\lnab_1\cdot\alv+m\gz+\halft V)i\dxi\, z_1\big] \nn\crt
\psi(\dxi\, z_2,\xv_2) &= U_{\mH}(\dxi\, z_2)\psi(0,\xv_2)U_{\mH}^\dag(\dxi\, z_2)
= \big[1+i\dxi\, z_2(i\rnab_2\cdot\alv-m\gz+\halft V)\big]\psi(0,\xv_2)
\end{align}
By symmetry, the potential at the quark position (due to the antiquark) is taken to be $\halft V(r)$ in the quark Hamiltonian, and analogously in the antiquark Hamiltonian\footnote{The equal time framework does not specify the potential for unequal time states. This assumption is motivated by continuity.}. The \order{\dxi} change in the potential due to the \order{\dxi} shifts in position is a negligible  \order{\dxi^2} correction. 

To simplify the notation I shall label the state using its momentum and wave function, \ie, in \eq{qcd97} $\ket{M,\Pv} \equiv \ketb{\Pv,\Phip}$.
Using \eq{freeh1} in \eq{infboost} gives after partial integrations
\begin{align} \label{freeh2} 
U_\mK(\dxi)\ketb{\Pv,\Phip} &= \Big|\Pv,\big[1+(i\rnab_1\cdot\alv-\halft\Pv\cdot\alv+m\gz+\halft V)i\dxi\, z_1\big]e^{i\dxi\alz/2}\Phip(\xv_1-\xv_2)e^{-i\dxi\alz/2} \nn\crt
&\times\big[1+i\dxi\, z_2(-i\lnab_2\cdot\alv+\halft\Pv\cdot\alv-m\gz+\halft V)\big]\Big\rangle
\end{align}
The derivatives acting on the coefficients $i\dxi\, z_1$ and $i\dxi\, z_2$ give,
\begin{align} \label{freeh3} 
1+i\rnab_1\cdot\alv(i\dxi\, z_1) = 1-\dxi\alz \simeq e^{-\dxi\alz} \hspace{2cm}
1-i(i\dxi\, z_2)\lnab_2\cdot\alv = 1+\dxi\alz \simeq e^{+\dxi\alz}
\end{align}
This reverses the sign in the exponents of \eq{freeh2}: $e^{\pm\dxi\alz/2}e^{\mp\dxi\alz}=e^{\mp\dxi\alz/2}$. 
Since the derivatives now operate only on $\Phi(\xv_1-\xv_2)$ we have $\lnab_2 \to -\lnab_1 \equiv -\lnab$, and it is convenient to define
\begin{align} \label{e3a2} 
\rhap \equiv i\alv\cdot\rnab-\halft \Pv\cdot\alv+m\gz \hspace{2cm}
\lhap \equiv -i\alv\cdot\lnab-\halft \Pv\cdot\alv+m\gz
\end{align}
The wave function $\Phip$ is assumed to satisfy the BSE \eq{qcd98},
\begin{align} \label{e3a3} 
\rhap\Phip(\xv) - \Phip(\xv)\lhap +V\Phip(\xv) = E\Phip(\xv)
\end{align}
where $\xv = \xv_1-\xv_2$ and $E=\sqrt{\Pv^2+M^2}$. We may express $z_1$ and $z_2$ as
\begin{align} \label{freeh4} 
z_1 = \halft(z_1+z_2)+ \halft z \hspace{2cm} z_2 = \halft(z_1+z_2)- \halft z  \hspace{2cm} z \equiv z_1-z_2
\end{align}
The term $\propto z_1+z_2$ in the wave function of \eq{freeh2} gives, using the BSE \eq{e3a3},
\begin{align} \label{freeh5} 
\Phip+\halft i\dxi(z_1+z_2)(\rhap\Phip-\Phip \lhap + V\Phip) \simeq e^{i\dxi E(z_1+z_2)/2} \Phip
\end{align}
This combines with the the plane wave factor in $\ket{\Pv,\Phip}$ \eq{qcd97} to $\exp[i(\Pv+\dxi E\hat\zv)\cdot(\xv_1+\xv_2)/2]$, where $\hat\zv$ is a unit vector in the $z$-direction. Hence the momentum of the boosted state is $\Pv+\dxi E\hat\zv$ as expected. Its wave function, \ie, the dependence on $\xv_1-\xv_2$, remains to be determined. 

The potential $V$ cancels in the term $\propto z_1-z_2=z$ of \eq{freeh2}. To first order in $\dxi$ we have, with $\Pv' \equiv \Pv+\dxi E\hat\zv$,
\begin{align} \label{freeh7} 
U_\mK(\dxi)\ketb{\Pv,\Phip}  &\equiv \ket{\Pv',\Phi^{(\Pv')}} = \ket{\Pv',\Phip} 
+ \ket{\Pv,\dxi\, \partial_\xi\Phip} \nn\crt
&= \Big\vert\Pv',\Phip\Big\rangle+\halft \dxi\Big\vert\Pv,iz (\rhap\Phip +\Phip\lhap) -\comb{\alz}{\Phip}\Big\rangle 
\end{align}
Comparing the two expressions gives
\begin{align} \label{ech2}
\Phi^{(\Pv')}(\xv) &\equiv \Phip(\xv)+\delta\xi\,\partial_\xi\Phip(\xv) \nn\crt
\partial_\xi\Phip &=\halft iz\big(\rhap\Phip +\Phip\lhap\big) -\halft\comb{\alz}{\Phip}
= \halft i\rhap\big(z\Phip\big) +\halft i\big(z\Phip\big)\lhap 
\end{align}
The BSE \eq{e3a3} implies (see Eq. 8.100 of \cite{Hoyer:2021adf}),
\begin{align} \label{ech2a} 
\rhap\Phip +\Phip\lhap=-\frac{2i}{E-V}\Pv\cdot\nv\Phip+\frac{i}{E-V}\comb{\alv\cdot\nv V}{\Phip}
\end{align}
Hence \eq{ech2} may be equivalently written, with $\Phip \equiv \Phip(\xv)$,
\begin{align} \label{ech3} 
\partial_\xi\Phip = \frac{z}{E-V}\,\Pv\cdot\nv\Phip - \frac{z}{2(E-V)}\,\comb{\alv\cdot\nv V}{\Phip} -\halft\comb{\alz}{\Phip}
\end{align}

%%%%%%%%%%%%%%%%%%%%%%%%%%
\subsection{The electromagnetic form factor} \label{sec3B}
%%%%%%%%%%%%%%%%%%%%%%%%%%

Gauges defined by a non-covariant condition such as $A^0=0$ generally change under boosts (see \cite{Dietrich:2012iy} for an example). Thus we may expect that $\Phi^{(\Pv')}$ given by \eq{ech2} differs by a gauge transformation of \order{\dxi} from the wave function which satisfies \eq{e3a3} with $\Pv \to \Pv+\dxi E\hat\zv$. However, the form factor \eq{f2} was previously shown to be gauge invariant \cite{Dietrich:2012un}. We can thus check the correctness of the boost by considering the frame dependence of the form factor. 

Eliminating $\rhap\Phip$ in $\partial_\xi\Phip$ \eq{ech2} using the BSE \eq{e3a3} gives
\begin{align} \label{ff2} 
\partial_\xi\Phip &= \Phip[\lhap+\halft(E-V)]iz -\halft\comb{\alz}{\Phip}
= \Phip\big[i\lnab\cdot\alv+\halft \Pv\cdot\alpha-m\gz-\halft(E-V)\big](-iz) -\halft\comb{\alz}{\Phip} \nn\crt
{\partial_\xi\Phip}^\dag &=iz\big[-i\rnab\cdot\alv+\halft\Pv\cdot\alv-m\gz-\halft(E-V)\big]{\Phip}^\dag+\halft\comb{\alz}{{\Phip}^\dag}
\end{align}
The EM current in \eq{f2} may be shifted to the origin using the generator of translations (four-momentum) $\hat P$,
\begin{align} \label{ff1}
j^\mu(y) = \bar\psi(y)\gamma^\mu\psi(y) = e^{i\hat P\cdot y}j^\mu(0)e^{-i\hat P\cdot y}
\end{align}
Since the $t=0$ bound states \eq{qcd97} are eigenstates of $\hat P$ the form factor and its Fourier transform become
\begin{align} \label{f2a}
F_{AB}^\mu(y) &= \bra{B, \Pvb}j^\mu(y)\ket{A,\Pva} = e^{i(P_B-P_A)\cdot y}\bra{B, \Pvb}j^\mu(0)\ket{A,\Pva}\nn\crt
F_{AB}^\mu(q) &= \int d^4y\,e^{-iq\cdot y}\,F^\mu_{AB}(y) \equiv (2\pi)^4\delta^4(P_B-P_A-q)G^\mu_{AB}(q)
\end{align}
The states $A$ and $B$ may be different, with no constraints on their masses $M$, momenta $\Pv$ or spins. We have explicitly,
\begin{align} \label{f3} 
G^\mu_{AB} = \int\Big(\prod_{i=1}^2 d\xv_i d\yv_i\Big)e^{i\Pva\cdot(\xv_1+\xv_2)/2-i\Pvb\cdot(\yv_1+\yv_2)/2} &\bra{0}\psi^\dag(0,\yv_2)\Phi_B^\dag(\yv_1-\yv_2)\gz\psi(0,\yv_1)\bar\psi(0,\bs0)\gamma^\mu\psi(0,\bs0) \nn\crt
&\times \bar\psi(0,\xv_1)\Phi_A(\xv_1-\xv_2)\psi(0,\xv_2)\ket{0}
\end{align}
The contractions $\{ \psi(0,\yv_1),\bar\psi(0,\bs0)\}= \gz\delta(\yv_1)$ and $\{ \psi(0,\bs0),\bar\psi(0,\xv_1)\}= \gz\delta(\xv_1)$ correspond to the virtual photon interacting with the fermion, whereas the antifermion is untouched, $\{ \psi^\dag(0,\yv_2),\psi(0,\xv_2)\}= \delta(\yv_2-\xv_2)$. For our present purposes it is sufficient to consider this contribution to the form factor. Denoting $\xv\equiv -\xv_2 =-\yv_2$ we get,
\begin{align} \label{f4} 
G^\mu_{AB} = \int d\xv\,e^{i(\Pvb-\Pva)\cdot\xv/2}\, \tr\big\{\Phi_B^\dag(\xv)\gamma^\mu\gz\Phi_A(\xv)\big\}
\end{align}

An infinitesimal boost in the $z$-direction transforms the three-momenta as $\partial_\xi\Pv=(0,0,E)$. Hence
\begin{align} \label{f5} 
\partial_\xi G^\mu_{AB} = \int d\xv\,e^{i(\Pvb-\Pva)\cdot\xv/2}\,\Big[(E_B-E_A)\halft iz\, \tr\big\{\Phi_A\Phi_B^\dag\gamma^\mu\gz\big\}+\partial_\xi\tr\big\{\Phi_A\Phi_B^\dag\gamma^\mu\gz\big\}\Big]
\end{align}
Using \eq{ff2} we find
\begin{align} \label{f6} 
\partial_\xi\tr\big\{\Phi_A\Phi_B^\dag\gamma^\mu\gz\big\}\Big]=\tr\big\{\big(&\Phi_A\big[i\lnab\cdot\alv+\halft\Pva\cdot\alv-m\gz-\halft(E_A-V)\big](-iz)-\halft\comb{\alz}{\Phi_A}\big)\Phi_B^\dag\gamma^\mu\gz \nn\crt
+&\Phi_A\big(iz\big[-i\alv\cdot\rnab+\halft \Pvb\cdot\alv-m\gz-\halft(E_B-V)\big]\Phi_B^\dag +\halft\comb{\alz}{\Phi_B^\dag}\big)\gamma^\mu\gz\big\}
\end{align}
The $m\gz$ and $V$ terms cancel in \eq{f6}, while the $E_A$ and $E_B$ contributions cancel the first term in \eq{f5}. The $\alv\cdot\nv$ terms combine into
\begin{align} \label{f7} 
\int d\xv\,e^{i(\Pvb-\Pva)\cdot\xv/2}\,&z\,\rnab\cdot\tr\big\{\Phi_A\alv\,\Phi_B^\dag\gamma^\mu\gz\big\} \nn\crt
%PHv2 Added final ]
=\int &d\xv\,e^{i(\Pvb-\Pva)\cdot\xv/2}\,\big[-\tr\big\{\Phi_A\alz\,\Phi_B^\dag\gamma^\mu\gz\big\}
-\halft iz\tr\big\{\Phi_A(\Pvb-\Pva)\cdot\alv\,\Phi_B^\dag\gamma^\mu\gz\big\} \big]
\end{align}
The second term cancels the $\Pva,\,\Pvb$ contributions to \eq{f6}. The commutators in \eq{f6} are
\begin{align} \label{f8} 
\halft\tr\big\{\big(\Phi_A\comb{\alz}{\Phi_B^\dag}-\comb{\alz}{\Phi_A}\Phi_B^\dag\big)\gamma^\mu\gz\big\}
= \tr\big\{\Phi_A\alz\,\Phi_B^\dag\gamma^\mu\gz\big\}-\halft\tr\big(\Phi_A\Phi_B^\dag\acomb{\alz}{\gamma^\mu\gz}\big)
\end{align}
The first terms of \eq{f7} and \eq{f8} cancel. The only remaining contribution is 
%PHv2 Added \big)
$-\halft\tr\big(\Phi_A\Phi_B^\dag\acomb{\alz}{\gamma^\mu\gz}\big)$. 
For $\mu=0$ we have $-\halft\acomb{\alz}{1}= \gamma^3\gz$, which gives $G^3_{AB}$. For $\mu=3$ similarly $-\halft\acomb{\alz}{\gamma^3\gz}= 1$, giving $G^0_{AB}$. For $\mu=1$ and $\mu=2$ the anticommutator vanishes. Thus,
\begin{align} \label{f9} 
\partial_\xi G^0_{AB} = G^3_{AB},\hspace{1cm} \partial_\xi G^3_{AB} = G^0_{AB},\hspace{1cm} \partial_\xi G^1_{AB} = \partial_\xi G^2_{AB}=0
\end{align}
showing that $G^\mu_{AB}$ transforms as a four-vector under infinitesimal boosts in the $z$-direction.
This is the first such demonstration for equal-time states, and supports the boost dependence \eq{ech2} of the wave function.

%%%%%%%%%%%%%%%%%%%%%%%%%%
\subsection{Bound state equation in temporal gauge} \label{sec3C}
%%%%%%%%%%%%%%%%%%%%%%%%%%

The gauge field $A^\mu$ transforms as a four-vector only up to a gauge transformation (\eg, section 8.1 of \cite{Weinberg:1995mt}). The gauge transformation required to maintain axial gauge under boosts in $D=1+1$ dimensions was demonstrated in \cite{Dietrich:2012iy}. Here the \order{\dxi} shifts of $\xv_1$ and $\xv_2$ in \eq{freeh1} can change the potential and gauge. Let us see whether the boosted wave function given by \eq{ech2} satisfies the (temporal gauge) bound state equation \eq{e3a3} with momentum $\Pv'=\Pv+\dxi E\hat\zv$ and energy $E'=E+\delta\xi\,P^z$, \ie, whether $BSE(\dxi) = \delta\xi\,\partial_\xi BSE(0)=0$, where
\begin{align} \label{e3c1} 
BSE(\dxi) \equiv (\rhap-\halft\dxi\,E\alz)\Phi^{(\Pv')}(\xv) - \Phi^{(\Pv')}(\xv)(\lhap-\halft\dxi\,E\alz)+\big(V'|\xv|-E-\dxi\,P^z\big)\Phi^{(\Pv')}(\xv) 
\end{align} 
Given that $BSE(0)=0$ and $\Phi^{(\Pv')}(\xv) = (1+\dxi\,\partial_\xi)\Phi^{(\Pv)}(\xv)$,
\begin{align} \label{e3c2} 
\partial_\xi BSE(0) =\rhap(\partial_\xi\Phip)-(\partial_\xi\Phip)\lhap-\halft E\comb{\alz}{\Phip}-(E-V)\partial_\xi\Phip
- P^z\Phip
\end{align}
The first two terms on the rhs. give, using the second expression for $\partial_\xi\Phip$ in \eq{ech2},
\begin{align} \label{e3c3} 
\halft i\rhap{\strut}^2(z\Phip)-\halft i(z\Phip)\lhap{\strut}^2 &= \halft i\big[(i\rnab-\halft \Pv)^2+m^2\big](z\Phip)- \halft i(z\Phip)\big[(i\lnab+\halft \Pv)^2+m^2\big] \nn\crt
&= \Pv\cdot\nv(z\Phip) = z\,\Pv\cdot\nv\Phip + P^z\Phip
\end{align}
From \eq{ech3} we have,
\begin{align} \label{e3c4} 
-(E-V)\partial_\xi\Phip =  \halft z\,\comb{\alv\cdot\nv V}{\Phip} +\halft(E-V)\comb{\alz}{\Phip} -z\,\Pv\cdot\nv\Phip
\end{align}
Using \eq{e3c3} and \eq{e3c4} in \eq{e3c2} we find
\begin{align} \label{e3c5} 
\partial_\xi BSE(0) = \halft z\,\comb{\alv\cdot\nv V}{\Phip} -\halft V\comb{\alz}{\Phip}
\end{align}
Thus $\partial_\xi BSE(0) \neq 0$ in general, showing that the boost must be combined with a gauge transformation to keep the state in temporal gauge. In section \ref{sec3E} below I determine the BSE that $\Phi^{(\Pv')}$ does satisfy, indicating the new gauge brought by the boost.

In the ``aligned'' quark configuration, where $\xv=\xv_1-\xv_2 =(0,0,z)$ is along the boost direction,  
\begin{align} \label{e3c6} 
z\,\alv\cdot\nv V'|z| = z\alz V'\veps(z) = V'|z|\alz = V\alz
\end{align}
Then $\partial_\xi BSE(0)=0$ in \eq{e3c5}, so the boost maintains temporal gauge. I next discuss some consequences of this.

%%%%%%%%%%%%%%%%%%%%%%%%%%
\subsection{The aligned configuration} \label{sec3D}
%%%%%%%%%%%%%%%%%%%%%%%%%%

Consider the BSE \eq{qcd98} with $\Pv=(0,0,P)$ and $\xv=(0,0,z)$ along the $z$-axis, so that $\partial_\xi BSE(0) =0$ in \eq{e3c5}. According to \eq{ech3},
\begin{align} \label{ech11} 
\partial_\xi\Phiz =  \frac{zP}{E-V}\,\dz\Phiz - \frac{E}{2(E-V)}\,\comb{\alz}{\Phiz} \hspace{2cm} (\xt=0)
\end{align}
where $E=M\cosh\xi$, $P=M\sinh\xi$ and $V=V'z$ (I take $z \geq 0$).
In the aligned configuration the boost implies the frame dependence (see section VIII.C.4 of \cite{Hoyer:2021adf}),
\begin{align}
\Phiz[z(\tau_P),\xtr=0] &= \exp(-\halft \zeta_P\alz)\,\Phi^{(0)}[z_0(\tau_0),\xtr=0]\exp(\halft \zeta_P\alz)  \label{qcd108aaa}\crt
\cosh\zeta_P(z) &\equiv \frac{E-V}{\sqrt{V'\tau_P(z)}} \hspace{2cm} \sinh\zeta_P(z) \equiv \frac{P}{\sqrt{V'\tau_P(z)}}  \label{qcd108bb}
\end{align}
The rest frame wave function $\Phi^{(0)}(z_0,\xt=0)$ is readily found using rotational symmetry as described in section \ref{sec2D}. The $z$-dependence of $\Phiz(z,\xt=0)$ is determined by the square of the kinematical momentum $\Pi=(E-V,\Pv)$,
\begin{align} \label{qcd106}
\tau_P(z) \equiv \big[(E-V)^2-P^2\big]/V' = (M^2-2EV+V^2)/V' 
\end{align}
To determine $\Phiz(z,\xtr=0)$ at some $z$ we need to find the corresponding rest frame coordinate $z_0$ of $\Phi^{(0)}(z_0,\xt=0)$ through the condition\footnote{Since $\tau_0(z_0) \geq 0$ this condition does not always give a real $z$. Here I assume $0 \leq z \leq (E-P)/V'$, for which $\tau_P(z) \geq 0$.}, 
\begin{align} \label{e3d1} 
\tau_P(z) = \tau_0(z_0) = (M-V'z_0)^2/V'
\end{align}

Let us verify\footnote{The derivation in \cite{Hoyer:2021adf} was based on the BSE, which required that \eq{qcd108aaa} hold also for its first transverse derivative. This is not generally true, as there can be \order{\xt} contributions to $\Phiz[z(\tau_P),\xtr]$. Here I derive the same result using the boost, which does not require a transverse derivative. Hence the relation \eq{qcd108aaa} (but not its $\partial_\perp$ derivative) always holds at $\xt=0$.} that the frame dependence \eq{qcd108aaa} of the wave function is consistent with \eq{ech11}. The rest frame wave function $\Phi^{(0)}(z_0,\xt=0)$ is frame independent if we take $\partial_\xi z_0 =0$. The frame independence of $\tau_P(z)$ in \eq{e3d1} then implies a correlated change of $\xi$ and $z$. The combination of partial derivatives $\partial_\xi$ and $\dz$ in \eq{ech11} (taken at fixed $z$ and $\xi$, respectively) indeed leaves $\tau_P(z)$ invariant,
\begin{align} \label{ech14} 
\Big(\partial_\xi - \frac{zP}{E-V}\,\dz\Big)V'\tau_P(z) = -2PV+\frac{2VP}{E-V}\,(E-V)=0
\end{align} 
The partial derivatives acting on $\zeta_P(z)$ (defined in \eq{qcd108bb}) give
\begin{align} \label{ech15} 
\partial_\xi\zeta_P(z) = \frac{M^2-EV}{V'\tau} \hspace{2cm} \partial_z\zeta_P(z) = \frac{P}{\tau}  \hspace{2cm} 
\Big(\partial_\xi - \frac{zP}{E-V}\,\dz\Big)\zeta_P(z) = \frac{E}{E-V}
\end{align}
Hence
\begin{align} \label{e3d2} 
\Big(\partial_\xi - \frac{zP}{E-V}\,\dz\Big)\exp(-\halft \zeta_P\alz)\,\Phi^{(0)}(\tau_0,\xt=0)\exp(\halft \zeta_P\alz) + \frac{E}{2(E-V)}\,\comb{\alz}{\Phiz} =0
\end{align}
which verifies the consistency of \eq{qcd108aaa} with \eq{ech11}.

The bound state mass $M$ was determined by the requirement of a regular wave function in the rest frame (section \ref{sec2D}).  The same mass should ensure the regularity of the wave function in any frame. Consider the $J^{PC}=0^{-+}$ state, whose rest frame wave function is given in \eq{qcd60}  (with $Y_{00}=1/\sqrt{4\pi}$). The explicit expression \eq{qcd108aaa} for $\Phiz(z,\xt=0)$ allows a check of regularity at $\xt=0$.

In the rest frame (with $z_0>0$, and absorbing $Y_{00}= 1/\sqrt{4\pi}$ into the normalization of $F_1$),
\begin{align}\label{e3d3}
\Phi^{(0)}(z_0,\xtr=0) &= \Big[\frac{2}{M-V'z_0}(i\alz\,\partial_{z_0}+m\gz)+1\Big]\gf\,F_1(z_0)
\end{align}
Using \eq{e3d1} gives $\partial_{z_0} = -2(M-V'z_0)\partial_{\tau_P}$, so that
\begin{align}\label{e3d4}
\Phiz(z,\xtr=0) = \Big(-4i\alz\,\partial_{\tau_P}+\frac{2m}{\sqrt{V'\tau_P}}\gz\,e^{\alz\zeta_P}+1\Big)\gf\,F_1\big[(M-\sqrt{V'\tau_P})/V'\big]
\end{align}
Rest frame regularity imposes $F_1(r=M/V')=0$, which with the radial equation \eq{qcd59} implies $\partial_rF_1(r=M/V')=0$. These conditions ensure the regularity of $\Phiz(z,\xtr=0)$ at $\tau_P=0$. 

The origin $(t=0,\xv=0)$ should be boost invariant, implying $z(z_0=0) = 0$. Then
$V'\tau_P(z=0)=V'\tau_0(z_0=0)=M^2$. According to \eq{qcd106} $V'\tau_P(z=0)=E^2-P^2$, imposing $E = \sqrt{P^2+M^2}$\,. 

At $z=z_0=0$ the potential vanishes and in \eq{qcd108bb} $\zeta_P(z=0)$ reduces to the standard boost parameter $\xi$. Then $\Phiz(\xv=0) = \exp(-\xi\alz/2)\Phir(\xv_0=0)\exp(\xi\alz/2)$ as for a free boost of Dirac spinors. This shows that the relative normalization of the rhs. and lhs. is correct in \eq{qcd108aaa}.

For weak potentials, $V \ll M$, we have $V'\tau_P \simeq M^2-2EV'z$ and $V'\tau_0 \simeq M^2-2MV'z_0$. Their equality \eq{e3d1} implies $z \simeq z_0\,M/E$, \ie, standard Lorentz contraction.

%%%%%%%%%%%%%%%%%%%%%%%%%%
\subsection{Bound state equation of the boosted state} \label{sec3E}
%%%%%%%%%%%%%%%%%%%%%%%%%%

In section \ref{sec3A} we saw \eq{freeh1} that an infinitesimal boost $\delta\xi$ in the $z$-direction shifts the fermion positions at $t=0$,
\begin{align} \label{boo1} 
\bar\psi(\delta\xi z_1,\xv_1) &= \bar\psi(0,\xv_1)\big[1+(-i\lnab_1\cdot\alv+m\gz+\halft V)i\delta\xi z_1\big] \nn\crt
\psi(\delta\xi z_2,\xv_2) &= \big[1+i\delta\xi z_2(i\rnab_2\cdot\alv-m\gz+\halft V)\big]\psi(0,\xv_2)
\end{align}
This was separated into an equal and an opposite shift: $z_{1,2}=\halft(z_1+z_2) \pm \halft(z_1-z_2)$. The equal shift $\propto z_1+z_2$ gave via the BSE rise to $\Pv \to \Pv'= \Pv+\dxi E\hat\zv$. In the opposite shift $\propto z_1-z_2$ the potential canceled at lowest order, and the \order{\delta\xi} change in $V$ could be ignored for the \order{\delta\xi} shift, giving the expressions \eq{ech2} and \eq{ech3} for $\partial_\xi\Phip(\xv)$.

In considering the \order{\delta\xi} BSE for the boosted wave function the change $\delta V = V' \delta |\xv_1-\xv_2|$ of the potential due to the shifts \eq{boo1} is relevant. Moreover, the boost should be combined with a gauge transformation in order to maintain temporal gauge. It is thus surprising that the BSE is nevertheless satisfied at \order{\delta\xi} when the quarks are aligned with the boost, $\xv = \xv_1-\xv_2 = (0,0,z)$ and the potential is linear, as seen from \eq{e3c5} and \eq{e3c6}.

It is instructive to determine the BSE satisfied by the boosted wave function for general $\xv$ when the gauge transformation and $\delta V$ are taken into account. The Lorentz structure of $V$ in the BSE \eq{qcd98} is the same as for $E$, \ie, it is the zeroth component of a 4-vector, $V(1,0,0,0)$. The boost turns this into $V(1,0,0,\delta\xi)$, generating an \order{\delta\xi} third component which appears as an $A^3$ field. With the quark and antiquark potentials each being $\halft V$ the $A^3$ potential adds $\delta\xi\,\halft V\comb{\alz}{\Phip}$ to the BSE \eq{e3c1}, which cancels the second term in \eq{e3c5}.

Since only opposite shifts change the BSE, let $z_1 \to \halft(z_1-z_2)\equiv \halft z$ and $z_2 \to -\halft z$ in \eq{boo1}. The quark is then shifted by $\bar\psi(0,\xv_1)\lnab_{1}\cdot\alv\halft z\,\delta\xi$. For the zeroth component of $\halft V(1,0,0,\delta\xi)$ to be boost invariant at \order{\delta\xi} the potential must counteract the shift, $\delta V_q = -\delta\xi\halft z\, \alv\cdot\nv V(\xv)$. The same reasoning for the antiquark gives $\delta V_{\bar q} = -\delta V_q$. Altogether we should then add $-\delta\xi\halft\,z \comb{\alv\cdot\nv V(\xv)}{\Phip}$ to the BSE \eq{e3c1}, which cancels the first term in \eq{e3c5}.

These arguments indicate that the boosted wave function satisfies the expected BSE at all $\xv$. The fact that the two terms in \eq{e3c5} cancel when $\xv=(0,0,z)$ appears here ``accidental''. In  $D=1+1$ dimensions this is required for the closure of the Lie algebra \cite{Dietrich:2012iy}. The terms added to the BSE \eq{e3c1} should be related to the gauge transformation which keeps the wave function in temporal gauge. I leave these considerations for further work.

%%%%%%%%%%%%%%%%%%%%%%%%%%
\section{General aspects of the $\Pv\neq 0$ wave functions} \label{sec4}
%%%%%%%%%%%%%%%%%%%%%%%%%%

In $D=1+1$ dimensions the BSE corresponding to \eq{qcd98} can be solved analytically, and the wave functions in different frames are explicitly related (see section VII A 4 of \cite{Hoyer:2021adf}). This holds for equal-time, color singlet states with a linear potential. No loop corrections are included, nor is the $N_C \to \infty$ limit taken as in the 't Hooft model \cite{tHooft:1974pnl}.

Solving \eq{qcd98} in $D=3+1$ dimensions is challenging since $\Pv$ breaks rotational symmetry. In this section I note some general aspects and then focus on a the simplest case of $J^{PC} = 0^{-+}$ states in section \ref{sec5}.

The BSE \eq{qcd98} remains invariant under rotations around $\Pv=(0,0,P)$, \ie, around the $z$-axis. The solutions may be characterized by their eigenvalue $\lm$ of $J^z = \halft \gf\alz -i(\xv\times\nv)^z$, by their parity $\eta_P$ and their charge conjugation $\eta_C$,
\begin{align}
\com{J^z}{\Phiz(\xv)}&=\lm\Phiz(\xv)  \label{e1.1a}\crt
\gz\Phiz(-\xv)\gz &=\eta_P\Phi^{(-P)}(\xv)  \label{e1.1b} \crt
\aly[\Phiz(-\xv)]^T\aly &=\eta_C\Phiz(\xv)  \label{e1.1c}
\end{align}
The symmetry under $J^z$ motivates using cylindrical coordinates $\xv=(z,\xt,\vphi)$, related to the cartesian $\xv=(x,y,z)$ as
\begin{align} \label{e4.1} 
z=z \hspace{1cm} x &= \xt\cos\vphi  \hspace{1cm} y = \xt\sin\vphi  \hspace{1cm} -\xv = (-z,\xt,\vphi+\pi)
\end{align}
Some useful relations are collected in the Appendix \ref{appA1}.

Instead of the rest frame expansion in \eq{qcd49} we may expand in the eigenfunctions $\exp(i\lm\vphi)$ of $L^z= -i\dphi$,
\begin{align} \label{e1.4}
\Phi_{\alpha\beta}^\lm(\xv) = \sum_{k=1}^{16}\Ga_{k}^{\alpha\beta}\,\phi_k^\lm(z,\xt)\,e^{i\lm\vphi}
\end{align}
The wave function $\Phi^\lm$ and its 16 components $\phi_k^\lm$ refer to the frame with CM momentum $\Pv=(0,0,P)$. A Dirac basis $\Ga_1 \ldots \Ga_{16}$ that is independent of $z$ and $\xt$ and commutes with $J^z=\halft \gf\alz-i\dphi$ is given by ($\alpha_i = \gz\gamma^i$)
\begin{align} \label{e1.4b}
  \begin{array}{ccccc}
    \Ga_1=1 & \Ga_2=i\alz & \Ga_3=i\at & \Ga_4=\ap & \crt 
    \Ga_5=i\gz & \Ga_6=i\gz\alz & \Ga_7=i\gz\at & \Ga_8=\gz\ap &\hspace{2cm}  \at \equiv \cos\vphi\, \alpha_1 + \sin\vphi\, \alpha_2\crt 
    \Ga_9=\gf & \Ga_{10}=i\gf\alz & \Ga_{11}=i\gf\at & \Ga_{12}=\gf\ap & \hspace{2.3cm} \ap \equiv - \sin\vphi\, \alpha_1+ \cos\vphi\, \alpha_2 \crt 
    \Ga_{13}=\gz\gf & \Ga_{14}=\gz\gf\alz & \Ga_{15}=\gz\gf\at & \Ga_{16}=i\gz\gf\ap  &
  \end{array}
\end{align}
Using the expansion \eq{e1.4} in the BSE \eq{qcd98} gives 16 relations between the components $\phi_k^\lm(z,\xt)$ of $\Phi^\lm(\xv)$. These are collected in the Appendix \ref{appA2}. %PHv2

Charge conjugation relates $\Phi^\lm(\xv)$ to $\Phi^\lm(-\xv)$ according to \eq{e1.1c}, whereas parity \eq{e1.1b} also reverses the CM momentum. Noting that $\at(\vphi+\pi)=-\at(\vphi)$ and similarly $\ap(\vphi+\pi)=-\ap(\vphi)$ we have
\begin{align} \label{e4.2}
  \begin{array}{llcl}
   & \eta^k_C = +1 &\mbox{for}& k=1,3,4,7,8,9,11,12,13,14 \\ 
\aly\big[\Ga_{k}(\vphi+\pi)\big]^T\aly = \eta^k_C\,\Ga_k(\vphi) \hspace{1cm} \\   
   & \eta^k_C = -1 &\mbox{for}& k=2,5,6,10,15,16
  \end{array}
\end{align}
Together with \eq{e1.1c} and \eq{e1.4} this gives
\begin{align} \label{e4.3} 
\phi_k^\lm(-z,\xt) = \eta_C\,\eta^k_C\,(-1)^\lm\, \phi_k^\lm(z,\xt)
\end{align}
It is thus sufficient to calculate the $\phi_k^\lm(z,\xt)$ for $z \geq 0$, with \eq{e4.3} giving the boundary condition at $z =0$.

%%%%%%%%%%%%%%%%%%%%%%%%%%
\section{Case study: The $J^{PC} = 0^{-+}$ state} \label{sec5}
%%%%%%%%%%%%%%%%%%%%%%%%%%

I illustrate the properties of the bound state wave functions using the $0^{-+}$ state. The rest frame $(P=0)$ wave function is given in \eq{qcd60}, where $F_1(r)$ satisfies the radial equation in \eq{qcd59} (with $j=0$). The expansions \eq{qcd57} and \eq{e1.4} define the relation between the $P=0$ wave functions\footnote{In the following I replace the $\lm=0$ superscript by the CM momentum: $\phi_k^{\lm=0}(z,\xt) \to \phip_k(z,\xt)$.} $\phi_k^{\scriptscriptstyle{(P=0)}}(z,\xt)$ and the radial functions $F_i(r)$,
\begin{align} \label{e1.26}
\phir_9(z,\xt)= F_1(r) \hspace{1cm} \phir_{10}(z,\xt)= zF_2(r) \hspace{1cm} \phir_{11}(z,\xt)= \xt F_2(r) \hspace{1cm} \phir_{13}(z,\xt)= F_4(r)
\end{align}
where $r= \sqrt{z^2+\xt^2}$\, and $Y_{00}=1/\sqrt{4\pi}$ was absorbed into the normalization of the radial functions.

%%%%%%%%%%%%%%%%%%%%%%%%%%
\subsection{Component equations} \label{sec5A}
%%%%%%%%%%%%%%%%%%%%%%%%%%

The BSE relations \eq{bsecomps} couple eight of the $J^{PC} = 0^{-+}$ component wave functions $\phip_k(z,\xt)$. Six of them may be expressed in terms of $\phip_8$ and $\phip_9$, reducing the system to two coupled partial differential equations (PDE's) of second order. With $V\equiv V'r$, $E=\sqrt{P^2+M^2}$ and $\tau_P$ of \eq{qcd106} now viewed as a function of $r= \sqrt{z^2+\xt^2}$\,,
\begin{align} \label{taudef} 
V'\tau_P(r) = (E-V)^2-P^2 = M^2-2EV+V^2
\end{align}
we get
\begin{align} \label{num10} 
\phip_4 &= \frac{2}{V'\tau_P}\big[P\dpp\phip_9+m(E-V)\phip_8\big] \nn\crt
\phip_{10} &= \frac{2}{E-V}\,\dz\phip_9 \nn\crt
\phip_{11} &= \frac{2}{V'\tau_P}\Big[(E-V)\dpp\phip_9+mP\phip_8\Big] \nn\crt
\phip_{13} &= \frac{2}{V'\tau_P}\Big[m(E-V)\phip_9+\frac{P}{\xt}\dpp\big(\xt\phip_8\big)\Big] \nn\crt
\phip_{14} &= \frac{2}{V'\tau_P}\Big[mP\phip_9+\frac{E-V}{\xt}\dpp\big(\xt\phip_8\big)\Big] \nn\crt
\phip_{15} &= -\frac{2}{E-V}\,\dz\phip_8 
\end{align}
The PDE's for $\phip_8$ and $\phip_9$ are coupled,
\begin{align} \label{num11} 
&\dpp\phip_{14}-\dz\phip_{15}-m\phip_4+\halft(E-V)\phip_8 =0 \nn\crt
&\inv{\xt}\dpp\big(\xt\phip_{11}\big)+\dz\phip_{10}-m\phip_{13}+\halft(E-V)\phip_9 =0
\end{align}
At $P=0$ these relations allow $\phir_k =0$ for $k=4,8,14,15$, and agree with \eq{e1.26} and \eq{qcd59} for $k=9,10,11,13$. For $m=0$ at general $P$ we have $\phip_k =0$ for $k=8,13,14,15$, and
\begin{align} \label{num12} 
&\phip_4 =\frac{2P}{V'\tau_P}\,\dpp\phip_{9} =\frac{P}{E-V}\,\phip_{11} \nn\crt
&\phip_{10} =\frac{2}{E-V}\,\dz\phip_{9} \nn\crt
&\phip_{11} =\frac{2(E-V)}{V'\tau_P}\,\dpp\phip_{9} \nn\crt
&\dz\phip_{10}+\inv{\xt}\dpp\big(\xt\phip_{11}\big)+\halft(E-V)\phip_9=0 \hspace{2cm} (m=0)
\end{align}

%%%%%%%%%%%%%%%%%%%%%%%%%%
\subsection{Boundary conditions} \label{sec5B}
%%%%%%%%%%%%%%%%%%%%%%%%%%

%%%%%%%%%%%%%%%%%%%%%%%%%%
\subsubsection{Boundaries at $z=0$ and $\xt=0$} \label{sec5B2}
%%%%%%%%%%%%%%%%%%%%%%%%%%

For a $J^{PC}=0^{-+}$ state \eq{e4.2} and \eq{e4.3} give
\begin{align} \label{e4.5} 
\phip_k(-z,\xt) &= +\phip_k(z,\xt) \ \ \ \mbox{for}\ k=4,8,9,11,13,14 \nn\crt
\phip_k(-z,\xt) &= -\phip_k(z,\xt) \ \ \ \mbox{for}\ k=10,15
\end{align}
The limit $z\to 0$ may be studied by assuming $\phip_8(z\to 0,\xt) \propto z^\alpha $ and $\phip_9(z\to 0,\xt) \propto z^\beta$. The terms of lowest power of $z$ in \eq{num11} arise from $\dz^2$ and are $\propto\alpha(\alpha-1)z^{\alpha-2}$ and $\propto\beta(\beta-1)z^{\beta-2}$, respectively. For these to vanish $\alpha$ and $\beta$ can take the values 0  or 1. The symmetries \eq{e4.5} fix $\alpha=\beta=0$.

Similarly, for $\xt\to 0$ at general $z$ we may assume $\phip_8(z,\xt\to 0) \propto \xt^\gamma $ and $\phip_9(z,\xt\to 0) \propto \xt^\delta$. The lowest powers $\xt^{\gamma-2}$ and $\xt^{\delta-2}$ in \eq{num11} arise from two transverse derivatives: $\dpp\phip_{14} \propto \dpp\big[\dpp(\xt\phip_8)/\xt\big] \propto (\gamma+1)(\gamma-1)\xt^{\gamma-2}$ and $\dpp(\xt\phip_{11})/\xt \propto \dpp(\xt\dpp\phip_{9})/\xt \propto \delta^2\xt^{\delta-2}$. The regular solutions have $\gamma=1$ and $\delta=0$ (exluding the singular $\log\xt$ behavior of $\phip_9$). This is consistent with the general requirement
\begin{align} \label{e4.6} 
\phip_k(z,\xt= 0) &=0 \ \ \ \mbox{for}\ k=4,8,11,15 
\end{align}
which ensures that the wave function is independent of $\vphi$ at $\xt=0$.
The above relations serve as boundary conditions at $(z=0,\xt)$ and $(z,\xt=0)$ for the PDE \eq{num10} and \eq{num11}.

%%%%%%%%%%%%%%%%%%%%%%%%%%
\subsubsection{Regions and singularities} \label{sec5B1}
%%%%%%%%%%%%%%%%%%%%%%%%%%

The denominators in \eq{num10} vanish at $V'\tau_P \equiv (V'r-E+P)(V'r-E-P)=0$ and at $V'r=E$, with $r=\sqrt{z^2+\xt^2}$. Local normalizability requires that the corresponding numerators also vanish. The regularity of the rest frame solutions $\phir_k(z,\xt)$ together with Poincar\'e invariance should ensure regularity at all $P$. I can here present only partial results. Regularity at $\xt=0$ was already verified in \eq{e3d4}. The Taylor expansion of the BSE at $r=0$ in section \ref{sec5C} below has a non-singular solution. In section \ref{sec5E} a numerical study of the $m=0$ wave function at $P=5\sqrt{V'}$ is consistent with regularity for $V'r \leq E-P$.

The function $\tau_P(r)$ \eq{taudef} can be negative when $P>0$. Based on the sign of $\tau_P$ we may distinguish three regions of $r$,
\begin{align} \label{e5.1a} 
&A: \hspace{1.3cm} 0\leq V'r\leq E-P \hspace{1cm} (\tau_P \geq 0) \nn\crt
&B: \hspace{.5cm} E-P \leq V'r\leq E+P  \hspace{1cm} (\tau_P \leq 0) \nn\crt
&C: \hspace{.5cm} E+P \leq V'r < \infty \hspace{1.7cm} (\tau_P \geq 0)
\end{align}
Region A is the standard ``valence quark'', contracting part of the wave function. Region C is dominated by the negative energy components of the quarks (see section VIII D 2 of \cite{Hoyer:2021adf}), tentatively associated with pair production (``string breaking''). The wave function is exponentially suppressed at large transverse coordinates $\xt$ in region B (see Appendix \ref{sec5D2}). This region grows with $P$ and may be related to tunneling: The production of pairs with CM momenta of \order{P} requires the potential energy $V'r$ to be of the same order.

%%%%%%%%%%%%%%%%%%%%%%%%%%
\subsubsection{Conditions at $r=0$, $V'r=E\mp P$ and $V'r=E$} \label{sec5B3}
%%%%%%%%%%%%%%%%%%%%%%%%%%

The constraint \eq{e4.6} requires $\phip_8$ to vanish at $r=0$, while the value of $\phip_9$ defines the the overall normalization,
\begin{align} \label{e5.1} 
\phip_8(z=0,\xt=0) = 0 \hspace{2cm} \phip_9(z=0,\xt=0) = 1
\end{align}

The regularity of $\phip_k(z,\xt)$ at $V'\tau_P= (V-E+P)(V-E-P)=0$ for $k=4,11,13,14$ in \eq{num10} requires,
\begin{align} \label{e4.7}
  \begin{array}{rc}
   \dpp\phip_9 \pm m\phip_8 = 0 \\ 
&  \hspace{1cm} \mbox{at}\ \ r = (E\mp P)/V' \\   
   \pm m\phip_9 +\inv{\xt}\dpp\big(\xt\phip_8\big) = 0 
  \end{array}
\end{align}
Similarly at $V'r = E$ the regularity of $\phip_{10}$ and $\phip_{15}$ requires
\begin{align} \label{e4.8} 
\dz\phip_{8} = \dz\phip_{9} = 0 \hspace{1.5cm} \mbox{at}\ \  r = E/V'
\end{align} 
 
%%%%%%%%%%%%%%%%%%%%%%%%%%
\subsection{Taylor expansion at $r=0$} \label{sec5C}
%%%%%%%%%%%%%%%%%%%%%%%%%%

The Taylor expansion of the rest frame radial function $F_1(r)$ in \eq{qcd57} and \eq{e1.26} is given by the radial equation \eq{qcd59},
\begin{align} \label{2.2.a} 
F_1(r)= 1-\frac{M^2-4m^2}{24}r^2+\frac{7M^2-4m^2}{144M}V'r^3+\frac{M^2(M^2-4m^2)^2-6V'^2(5M^2+4m^2)}{1920M^2}r^4 +\morder{r^5}
\end{align}
The neglected \order{r^5} terms leave a residue of \order{r^3} in the radial equation. The \order{r^2} term in \eq{2.2.a} is independent of $V'$ because the radial equation is independent of the potential for $V'r \ll M$.

For $P>0$ it is convenient to switch from cylindrical to spherical coordinates, $\xv=r(\sin\theta\cos\vphi,\sin\theta\sin\vphi,\cos\theta)$. Instead of the polar angle $\theta$ I shall use $t \equiv \cos\theta = z/r$, with $0\leq t \leq 1$. The derivatives are related as in \eq{eA.2}. The wave functions are obviously different functions of the spherical coordinates, but will for conciseness nevertheless be similarly denoted $\phip_k(r,t)$. The arguments will be clear from the context, or written out explicitly.

In accordance with \eq{e4.6} and \eq{e5.1} I express the power series in $r$ for $\phip_8$ and $\phip_9$ as
\begin{align} \label{2.2.1} 
\phip_8(r,t) = \sqrt{1-t^2}\;\sum_{j=1}^2 h_{8j}(t)r^j + \morder{r^3} \hspace{3cm}
\phip_9(r,t) = 1+\sum_{j=1}^2 h_{9j}(t)r^j  + \morder{r^3}
\end{align}
The remaining six components $\phip_k$ are given in terms of these by \eq{num10}. The BSE conditions \eq{num11} determine second order differential equations for the coefficient functions $h_{8j}(t)$ and $h_{9j}(t)$, which may be solved analytically. Imposing $t \to -t$ symmetry \eq{e4.5} and regularity for $t \to 1\ (\xt\to 0)$ the BSE at \order{r^0} implies
\begin{align} \label{2.2.3} 
\phip_8(r,t) &=\sqrt{1-t^2}\; \frac{2mV'E}{P}\bigg\{\frac{-1}{3M^2}\Big[E+\frac{E^2-(3E^2-2P^2)(1-t^2)}{E+\sqrt{M^2+P^2t^2}}\Big]+\frac{E\,t}{P^2}\log\Big[\frac{E\,t+\sqrt{M^2+P^2t^2}}{M(1+t)}\Big]\bigg\}r^2 +\morder{r^3} \nn\crt
\phip_9(r,t) &= 1-\frac{(M^2-4m^2)(M^2+P^2t^2)}{24M^2}r^2+\morder{r^3} = 1-\frac{M^2-4m^2}{24}\Big(\frac{E^2}{M^2}\,z^2+\xt^2\Big) +\morder{r^3}
\end{align}
where $E = \sqrt{M^2+P^2}$. $\phip_9$ is at \order{r^2} independent of the potential, and given by the radial function $F_1(r)$ \eq{2.2.a} with a Lorentz contracted $z$-coordinate. On the other hand, $\phip_8 \propto V'$ arises only due to the interaction. It vanishes in the rest frame as well as for $m=0$,
\begin{align} \label{e5.2} 
\phip_8(r,t) = -\sqrt{1-t^2}\; \frac{mV'P}{2M^2}r^2 +\morder{P^2,\,r^3}
\end{align}
In the $t \to 1$ limit,
\begin{align} \label{2.2.6} 
\phip_8(r,t) = &-\sqrt{1-t^2}\; \frac{mV'}{M^2 P^3}\bigg\{E^2\Big[P^2-2M^2\log\Big(\frac{E}{M}\Big)\Big]-\frac{1-t}{2}\Big[P^2(E^2+M^2)-4M^2E^2\log\Big(\frac{E}{M}\Big)\Big]\bigg\}r^2 \nn\crt
&+\morder{(1-t)^{5/2},\,r^3}
\end{align}
The corresponding Taylor expansions of the $\phip_k$ in \eq{num10} are, with $\mu \equiv \sqrt{M^2+P^2t^2}$,
\begin{align} \label{2.2.8} 
\phip_4(r,t) &= -\sqrt{1-t^2}(M^2-4m^2)\frac{P}{6M^2}\,r + \morder{r^2} \nn\crt
\phip_{10}(r,t) &= -t(M^2-4m^2)\frac{E}{6M^2}\,r + \morder{r^2} \nn\crt
\phip_{11}(r,t) &= -\sqrt{1-t^2}(M^2-4m^2)\frac{E}{6M^2}\,r + \morder{r^2} \nn\crt
\phip_{13}(r,t) &= \frac{2mE}{M^2} -\frac{2mV'}{M^2P^2}\Big\{P^2+4E\mu-4E^2\Big[1+t\log\Big(\frac{Et+\mu}{M(1+t)}\Big)\Big]\Big\}\,r + \morder{r^2} \nn\crt
\phip_{14}(r,t) &= \frac{2mP}{M^2} -\frac{4mEV'}{M^2P^3}\Big\{P^2+2E\mu-2E^2\Big[1+t\log\Big(\frac{Et+\mu}{M(1+t)}\Big)\Big]\Big\}\,r + \morder{r^2} \nn\crt
\phip_{15}(r,t) &= \frac{4mEV'}{M^2P^3\sqrt{1-t^2}}\Big\{Et(E-\mu)-M^2(1-t^2)\log\Big(\frac{Et+\mu}{M(1+t)}\Big)\Big\}\,r + \morder{r^2}
\end{align}

The behavior of the wave functions at large $r$ is considered in Appendices \ref{sec5D} and \ref{sec5D2}.

%%%%%%%%%%%%%%%%%%%%%%%%%%
\subsection{Numerical study of the ground $J^{PC}=0^{-+}$ state for $0\leq r \leq E-V$ and $m=0$} \label{sec5E}
%%%%%%%%%%%%%%%%%%%%%%%%%%

The wave function of any state at \order{\as^0} is found by solving the Bound State Equation (BSE) \eq{qcd98} with the linear potential \eq{2b4}. The Dirac expansion \eq{e1.4} gives 16 coupled Partial Differential Equations (PDA's) \eq{bsecomps}. The $J^{PC}=0^{-+}$ states have the boundary conditions of section \ref{sec5B2}, as well as the conditions of regularity in section \ref{sec5B3}. In the rest ($\Pv=0$) frame the mass $M$ of the bound state is determined by regularity at $r=0$ and $r=M/V'$.

Poincar\'e symmetry should ensure the regularity of the wave function for $\Pv\neq 0$ and energy $E=\sqrt{M^2+\Pv^2}$. According to \eq{e1.26} the radial function $F_1(r)$ of \eq{qcd57} equals the rest frame $\gf$ component $\phir_9$, so \eq{qcd108aaa} implies,
\begin{align} \label{e5.11} 
\phip_9(z,\xt=0) = F_1\big[(M-\sqrt{(M-V'z)^2-P^2}\,)/V'\big]
\end{align}
where $\Pv=(0,0,P)$. This ensures \eq{e3d4} the regularity of the full wave function $\Phip(z,\xt=0)$.

To test the regularity of $\Phiz(\xv)$ for $\xt>0$ I numerically solve the $J^{PC}=0^{-+}$ ground state wave function. Choosing quark mass $m=0$ sets $\phip_k =0$ for $k=8,13,14,15$, simplifying the PDE's to \eq{num12}. The range $0 \leq r \leq (E-P)/V'$ allows to verify the regularity condition \eq{e4.7}, $\partial_\perp\phip_9(z,\xt)=0$ at $r = (E-P)/V'$.

%%%%%%%%%%%%%%%%%%%%%%%%%%
\subsubsection{Rest frame: Radial function $F_1(r)$} \label{sec5E1}
%%%%%%%%%%%%%%%%%%%%%%%%%%

The radial function $F_1(r)$ satisfies \eq{qcd59}, which for a $J^{PC}=0^{-+}$ state with $m=0$ becomes,
\begin{align} \label{e5.12}
F_1''(r)+\Big(\frac{2}{r}+\frac{V'}{M-V'r}\Big)F_1'(r)+ \quart (M-V'r)^2F_1(r) = 0
\end{align}
I expand $F_1(r)$ in terms of Chebyshev polynomials\footnote{Some relations for the Chebyshev polynomials are given in Appendix \ref{appB1}.} $T_n(2V'r/M-1)$, whose argument spans the range $[-1,1]$ for $0\leq r \leq M/V'$ \cite{Boyd:2000jp}. The parametrization imposes the boundary conditions $F_1(r=0)=1$ and $F_1(r=M/V')=0$:
\begin{align} \label{e5.13} 
F_1(r)= 1-\frac{V'r}{M}+\sum_{n=1}^{n_0}\Big\{c_{2n}\Big[T_{2n}\Big(\frac{2V'r}{M}-1\Big)-1\Big] + c_{2n+1}\Big[T_{2n+1}\Big(\frac{2V'r}{M}-1\Big)-\frac{2V'r}{M}+1\Big]\Big\}
\end{align}
The parameters $M$ and $c_n\ (n=2,\ldots,2n_0+1)$ are fit\footnote{I used the Mathematica program FindFit.} to the ``data'' that the lhs. of the radial equation \eq{e5.12} vanish at $n_p$ values of $r$, evenly distributed in the $0\leq r \leq M/V'$ interval. Using $n_0=10$ and $n_p=500$ gave
\begin{align} \label{e5.14} 
M=3.79588\,\sqrt{V'}
\end{align}
and parameters $c_n$ such that $F_1(r)$ \eq{e5.13} satisfies the radial equation \eq{e5.12} at \order{10^{-11}}.

%%%%%%%%%%%%%%%%%%%%%%%%%%
\subsubsection{Wave function at $P=5\,\sqrt{V'}$ \label{sec5E2}}
%%%%%%%%%%%%%%%%%%%%%%%%%%

I numerically solve the BSE \eq{num12} for the $\phip_k(r,t=z/r)$ at $P=5\,\sqrt{V'}$ for $0\leq r \leq (E-P)/V' = 1.27746/\sqrt{V'}$\,, \ie, in Region $A$ of \eq{e5.1a}. The derivatives $\dz$ and $\dpp$ are expressed in terms of $\dr$ and $\dt$ in \eq{eA.2}. 

According to \eq{e4.5} $\phip_9(r,t)$ is symmetric under $t \to -t$ and is thus expanded in even Chebyshev polynomials $T_{2\ell}(t)$.  $\phip_9(r=0,t)=1$ \eq{2.2.3} and $\phip_9(r,t=1)$ is given by the rest frame radial function $F_1$ \eq{e5.11}. This implies the Taylor expansion of $\phip_9(r,t=1)$ at $r=(E-P)/V'$,
\begin{align} \label{e5.15} 
\phip_9(r,t=1) = F_1''(M/V')\bigg[\frac{P}{V'}\Big(\frac{E-P}{V'}-r\Big) + \frac{4\sqrt{2}P^{3/2}}{3M\sqrt{V'}}\Big(\frac{E-P}{V'}-r\Big)^{3/2} + \morder{\Big(\frac{E-P}{V'}-r\Big)^{2}}\bigg]
\end{align}
where $F_1''(M/V') = 0.0572986\,V'$. We can anticipate that $\phip_9$ will not be accurately described by low order polynomials close to its branch point at $r=(E-P)/V',\ t=1$.

I used the parametrization,
\begin{align} \label{e5.16} 
\phip_9(r,t) = 1 + \sum_{n=1}^{n_m}\sum_{\ell=0}^{\ell_m}a_{n,2\ell}\Big[T_{n}\Big(\frac{2V'r}{E-P}-1\Big)-(-1)^n\Big]T_{2\ell}(t)
\end{align}
The first five coefficients $a_{n,0}$ were constrained by imposing \eq{e5.11} at five values $r=r_j$, 
\begin{align} \label{e5.17} 
\phip_9(r_j,t=1)= F_1\big[\big(M-\sqrt{(E-V'r_j)^2-P^2}\,\big)/V'\big]\ \ \ \mbox{for}\ \ \ r_j = (E-P)j/5V'\ \ (j=1,\ldots,5)
\end{align}
The remaining $a_{n,2\ell}$ were determined by a fit to the BSE \eq{num12}, similarly as for $F_1(r)$ in \eq{e5.13}.

\begin{figure}[h] \centering
\includegraphics[width=.6\columnwidth]{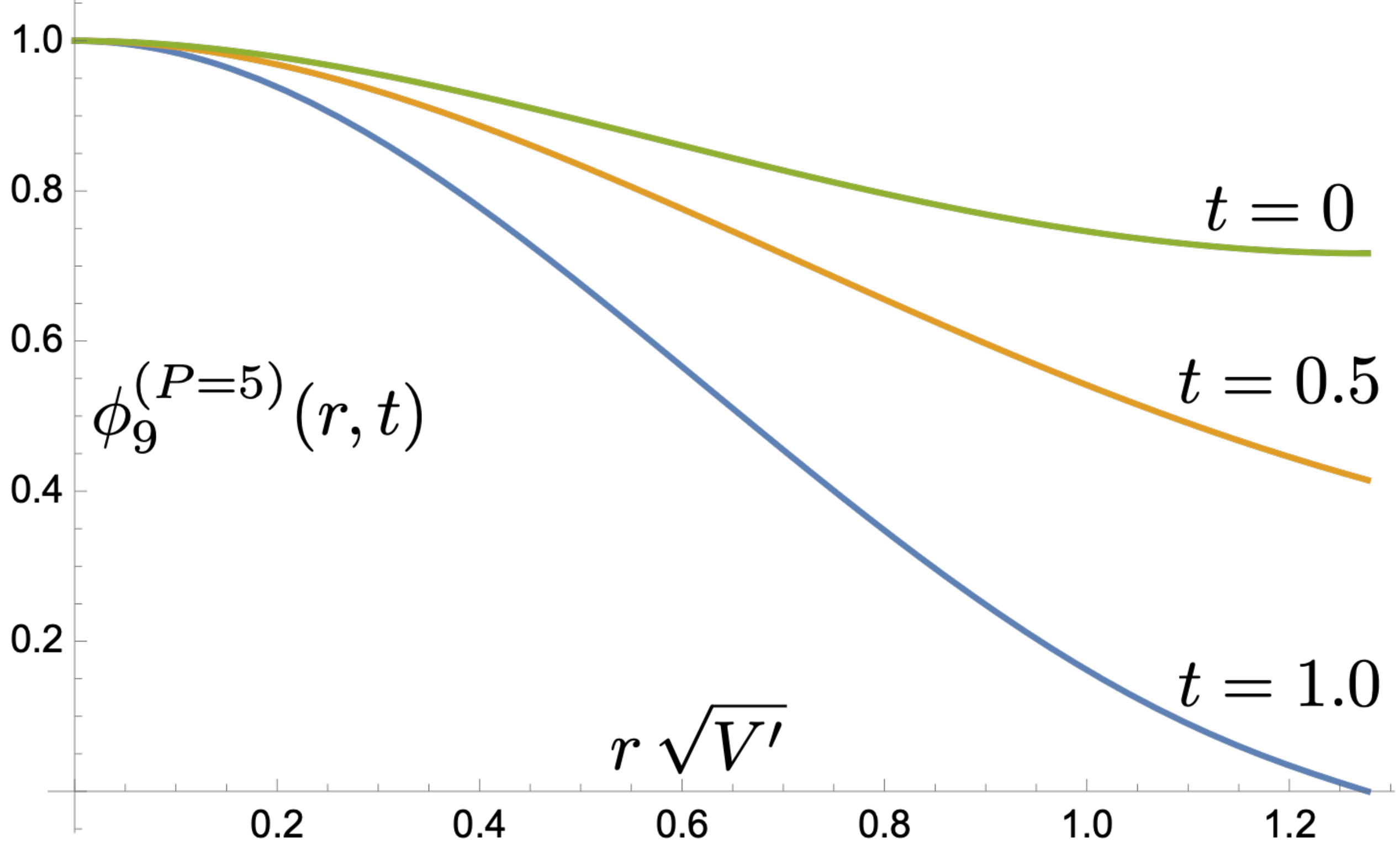}
\caption{The wave function $\phip_9(r,t)$ at $P=5\,\sqrt{V'}$ shown for $0 \leq r \leq (E-P)/V'$ at three values of $t=z/r$. \label{phi9}}
\end{figure}

With $n_m=20,\ n_\ell=10$ and 100 equidistant data points in $0 \leq r \leq (E-P)/V'$ and 50 points in
$0 \leq t \leq 1$ the fit returned parameters $a_{n,2\ell}$ for which the parametrization \eq{e5.16} satisfied the BSE at \order{10^{-8}}. The relation \eq{e5.11} was fulfilled at \order{10^{-3}}, with a maximum deviation of $3\cdot 10^{-3}$ close to the branch point $r=(E-P)/V'$.  The regularity condition \eq{e4.7}, $\dpp\phip_9(r=(E-P)/V',t)=0$ was fulfilled at \order{10^{-11}} for all values of $t$. Plots of $\phip_9(r,t)$ are shown in \fig{phi9}.

The quality of the fit indicates that there is a regular wave function which satisfies the BSE \eq{num12} in region $A$ of \eq{e5.1a}. The extension to regions $B$ and $C$, as well as solutions with $m \neq 0$ and for states with other quantum numbers, is left for future work.

%%%%%%%%%%%%%%%%%%%%%%%%%%%%%%%%%%%%%%%%%%%%%%%%%%%%%%%%
%%%%%%%%%%%%%%%%%%%%%%%%%%%%%%%%%%%%%%%%%%%%%%%%%%%%%%%%%

%%%%%%%%%%%%%%
\section{Summary  \label{sec6}}
%%%%%%%%%%%%%%
 
Poincar\'e covariance for bound states is dynamic (involves interactions). Any formally exact framework should have Poincar\'e symmetry, and this requires that the structure of the theory is maintained. QCD poses a special challenge since its confinement scale $\lqcd$ is not a parameter of the action. I have argued that $\lqcd$ may be introduced via a boundary condition, which leaves the equations of motion intact. To test the specific approach \cite{Hoyer:2021adf} I here studied the frame dependence of the bound states and their gauge invariant form factors.

Atoms are bound by the classical Coulomb potential $-\alpha/r$, to which higher order quantum corrections are added perturbatively. Heavy quarkonia are well described in a similar approach, when a phenomenological linear term $\la^2r$ is added to the Coulomb potential. The ``Cornell potential'' \cite{Eichten:1979ms,Eichten:2007qx} first determined by quarkonium data agrees with lattice calculations. Confinement may then be described by a classical field, in analogy to atoms.

The $q\bar q$ and $qqq$ quantum numbers of relativistically bound hadrons indicates that the potential is \textit{instantaneous in time}. The gauge-dependent $A^0$ and longitudinal $\Av_L$ fields generate instantaneous potentials when the gauge fixing condition is independent of time. The Coulomb ($\nv\cdot\Av=0$) and temporal ($A^0=0$) gauges are primary alternatives since they preserve explicit (kinetic) translation and rotation symmetry.

In Coulomb gauge Gauss' law $G \equiv \delta S/\delta A^0=0$ is an operator equation of motion, which expresses $A^0$ in terms of the dynamical (propagating) quark and gluon fields. The absence of the conjugate field $\partial_t A^0$ in the action requires to implement canonical quantization with constraints \cite{Christ:1980ku,Weinberg:1995mt}.

In temporal gauge Gauss' law ($G=0$) is not an equation of motion, and the electric field $\Ev_L$ is conjugate to $\Av_L$. However, the $A^0=0$ condition does not fully fix the gauge. It allows time-independent gauge transformations, which are generated by Gauss' operator $G$. Gauss' law is implemented as a \textit{constraint on physical states,} $G\ket{phys}=0$, ensuring the full gauge fixing of $\ket{phys}$ \cite{Willemsen:1977fr,Bjorken:1979hv,Leibbrandt:1987qv,Strocchi:2013awa}. Gauss' constraint determines, for each physical state, a classical longitudinal electric field $\Ev_L$, to which quantum corrections can then be added.

States such as $\sum_A\ketb{q_\alpha^A(t,\xv_1)\,\bar q_\beta^A(t,\xv_2)}$, which are singlets under global color transformations, have a vanishing classical electric field, $\Ev_L^a(t,\xv)=0$ for all $\xv$. This is specific to non-abelian theories, as it results from the sum over quark colors $A$. In QED an $\ketb{e_\alpha^-(t,\xv_1)\,e_\beta^+(t,\xv_2)}$ state does have a classical electric dipole field. 

In \cite{Hoyer:2021adf} I introduced a new boundary condition to the gauge constraint $G\,\ketb{phys}=0$. This involves a scale $\la$ and gives rise to a confining instantaneous potential. We may then set $\as=0$ and consider whether the hadron dynamics at \order{\as^0} can serve as the lowest order term of a perturbative expansion. The resulting bound states have some intriguing and promising properties discussed in \cite{Hoyer:2021adf}. Here I focussed on their boost covariance. 

The bound state momentum $\Pv \neq 0$ prevents the separation of radial and angular variables in the bound state equation (BSE), which must then be solved as a set of coupled partial differential equations. Boost covariance requires that the wave function, satisfying specific boundary conditions and having eigenvalue $E=\sqrt{M^2+\Pv^2}$, is everywhere regular (locally normalizable). This is not evident from the BSE. However, in all cases I studied the requirements of Poincar\'e covariance turned out to be fulfilled.

A boost transforms states with constituents at equal time into unequal-time states. Time equality can be restored through time translations of each constituent. The boost changes the gauge and shifts the spatial coordinates of the constituents by unequal amounts.

Remarkably, the (transition) electromagnetic form factors of $q\bar q$ states with any mass, spin and momentum transform covariantly under infinitesimal boosts. Hence physical (gauge invariant) quantities might be evaluated using states in gauges which transform simply under boosts \cite{Hoyer:2016aew}. 

The present results suggest that the \order{\as^0} sector can serve as the lowest order term in a ``Bound Fock Expansion'' \cite{Hoyer:2021adf}. This would open up hadron physics to perturbative analyses, with the terms of higher orders in $\as$ being determined as in \eq{e2b2}.

\begin{acknowledgments}
%PHv2 Added s
I thank Matti J\"arvinen for helpful discussions.
\end{acknowledgments}
 
\appendix
\renewcommand{\theequation}{\thesection.\arabic{equation}}

%%%%%%%%%%%%%%
\section{Expressions related to section \ref{sec4}} \label{appA}
%%%%%%%%%%%%%%

%%%%%%%%%%%%%%
\subsection{Coordinates and Dirac matrices \label{appA1}}
%%%%%%%%%%%%%%

The cylindrical coordinates $\xv=(z,\xt,\vphi)$ are related to the cartesian $\xv=(x,y,x)$ as
\begin{align} \label{eA.1}
z=z \hspace{1cm} x &= \xt\cos\vphi  \hspace{1cm} y = \xt\sin\vphi  \hspace{1cm} -\xv = (-z,\xt,\vphi+\pi) \nn\crt
\partial_x &= \cos\vphi\,\dpp-\inv{\xt}\sin\vphi\,\dphi \hspace{1cm} 
\partial_y = \sin\vphi\,\dpp+\inv{\xt}\cos\vphi\,\dphi \nn \crt
\dpp &= \cos\vphi\,\partial_x+\sin\vphi\,\partial_y \hspace{1.1cm} 
\inv{\xt}\dphi = -\sin\vphi\,\partial_x+\cos\vphi\,\partial_y  \hspace{1cm}
\end{align}
The spherical coordinates $\xv=(r,t,\vphi)$ are related to the cylindrical ones as
\begin{align} \label{eA.2} 
r &= \sqrt{z^2+\xt^2} \hspace{4.5cm} t \equiv \cos\theta = \frac{z}{r}  \hspace{2cm} \vphi=\vphi \nn\crt
\dz &= t\,\dr +\frac{1-t^2}{r}\,\dt  \hspace{3.5cm} 
\dpp = \sqrt{1-t^2}\Big(\dr -\frac{t}{r}\,\dt\Big) \nn\crt
\dr &= \inv{\sqrt{z^2+\xt^2}}\big(z\dz+\xt\dpp\big) \hspace{2cm} 
\dt = \sqrt{z^2+\xt^2}\Big(\dz -\frac{t}{\sqrt{1-t^2}}\dpp\Big)
\end{align} 

The following relations are useful:
\begin{align} \label{eA.3} 
 \at &\equiv \cos\vphi\, \alpha_1 + \sin\vphi\, \alpha_2 \hspace{2.7cm} \ap \equiv - \sin\vphi\, \alpha_1+ \cos\vphi\, \alpha_2 \nn \crt
\xv_\perp\cdot\alv_\perp &= x \alpha_1+y \alpha_2 = \xt\,\at \hspace{1.1cm} (\xv_\perp\times\alv_\perp)^z = x \alpha_2-y \alpha_1 = \xt\,\ap \nn\crt
\acom{\at}{\at} &= \acom{\ap}{\ap} =2 \hspace{2.5cm} \acom{\alz}{\at} = \acom{\alz}{\ap} = \acom{\at}{\ap} =0 \nn\crt
\com{\alz}{\at} &= 2i\ap\gf \hspace{1cm} \com{\alz}{\ap} = -2i\at\gf \hspace{1cm}  \com{\at}{\ap} = 2i\alz\gf \nn\crt
\com{\dpp}{\at} &= \com{\dpp}{\ap} = 0 \hspace{1cm} \com{\dphi}{\at} =\ap \hspace{1cm} \com{\dphi}{\ap} =-\at \nn\crt
\alpha_1\partial_x+\alpha_2\partial_y &= \at\rder_\perp+\inv{\xt}\ap\rder_\vphi = \big(\rder_\perp+\inv{\xt}\big)\at + \inv{\xt}\rder_\vphi\ap = \lder_\perp\at + \lder_\vphi \inv{\xt}\ap
\end{align}

%%%%%%%%%%%%%%
\subsection{Relations between the 16 component wave functions  \label{appA2}}
%%%%%%%%%%%%%%

Using the expansion \eq{e1.4} in the BSE \eq{qcd98} with $\Pv = (0,0,P)$ gives the following 16 relations between the components $\phi_k \equiv \phi_k^\lm(z,\xt)$ of $\Phi^\lm(\xv)$:

\begin{subequations} \label{bsecomps}
\begin{align}
-\partial_z\phi_{2}-\partial_\perp\phi_{3}-\inv{\xt}(\phi_{3}+\lm\phi_{4}) &= \halft(E-V)\phi_{1}  \label{e1.19a} \crt
\partial_z\phi_1+m\phi_6 &= \halft(E-V)\phi_2 \label{e1.19b}\crt
\partial_\perp\phi_1+\frac{P}{2}\phi_{12}+m\phi_7 &= \halft(E-V)\phi_3  \label{e1.19c} \crt
-\frac{\lm}{\xt}\phi_{1}+\frac{P}{2}\phi_{11}+m\phi_{8} &= \halft(E-V)\phi_{4}  \label{e1.19d} \crt
\frac{P}{2}\phi_{6} &= \halft(E-V)\phi_{5}  \label{e1.19e} \crt
\partial_\perp\phi_{16}+\inv{\xt}(\phi_{16}-\lm\phi_{15})+\frac{P}{2}\phi_{5}+m\phi_{2} &= \halft(E-V)\phi_{6}  \label{e1.19f} \crt
-\partial_z\phi_{16}+\frac{\lm}{\xt}\phi_{14}+m\phi_{3} &= \halft(E-V)\phi_{7}  \label{e1.19g} \crt
\partial_z\phi_{15}-\partial_\perp\phi_{14}+m\phi_{4} &= \halft(E-V)\phi_{8}  \label{e1.19h} \crt
-\partial_z\phi_{10}-\partial_\perp\phi_{11}-\inv{\xt}(\phi_{11}+\lm\phi_{12})+m\phi_{13} &= \halft(E-V)\phi_{9}  \label{e1.19i} \crt
\partial_z\phi_{9} &= \halft(E-V)\phi_{10}  \label{e1.19j} \crt
\partial_\perp\phi_{9}+\frac{P}{2}\phi_{4} &= \halft(E-V)\phi_{11}  \label{e1.19k} \crt
-\frac{\lm}{\xt}\phi_{9}+\frac{P}{2}\phi_{3} &= \halft(E-V)\phi_{12}  \label{e1.19l} \crt
\frac{P}{2}\phi_{14}+m\phi_{9} &= \halft(E-V)\phi_{13}  \label{e1.19m} \crt
\partial_\perp\phi_{8}+\inv{\xt}(\phi_{8}+\lm\phi_{7})+\frac{P}{2}\phi_{13} &= \halft(E-V)\phi_{14}  \label{e1.19n} \crt
-\partial_z\phi_{8}-\frac{\lm}{\xt}\phi_{6} &= \halft(E-V)\phi_{15}  \label{e1.19o} \crt
\partial_z\phi_{7}-\partial_\perp\phi_{6} &= \halft(E-V)\phi_{16}  \label{e1.19p}
\end{align}
\end{subequations}

%%%%%%%%%%%%%%
\section{Expressions related to section \ref{sec5}} \label{appB}
%%%%%%%%%%%%%%

%%%%%%%%%%%%%%%%%%%%%%%%%%
\subsection{Asymptotic behavior of the $J^{PC} = 0^{-+}$ wave function for $r \to \infty$ at fixed $P$} \label{sec5D}
%%%%%%%%%%%%%%%%%%%%%%%%%%

The leading behavior of the radial function $F_1(r)$ in the limit of $r \to\infty$ is given in \eq{qcd114}. The term $\propto \Pv$ in the BSE \eq{qcd98}, which breaks rotational invariance, is suppressed by one power of $r$ compared to the potential term $\propto V'r$. Hence we may expect rotational symmetry to be restored at large $r$. Assuming $\dr \propto r \gg \dt$ the BSE \eq{num10} and \eq{num11} indeed reduce at leading order in $r$ to
\begin{align} \label{e5.3} 
\dr^2\phip_k(r,t) + \quart V^2\phip_k(r,t)=0 \hspace{2cm} (k=8,9;\ \ r \to \infty)
\end{align} 
Thus the bound state conditions for $\phip_8(r,t)$ and $\phip_9(r,t)$ decouple at large $r$, and agree with that for $F_1(r)$ in the radial equation \eq{qcd59}. With $\tau_P(r)$ given by \eq{taudef} the expressions
\begin{align} \label{e5.4} 
\phip_8(r,t) &= N_8\sqrt{1-t^2}\,r^{-1-im^2}\exp\big[\quart\tau_P(r)\big] \nn\crt
\phip_9(r,t) &= N_9\,r^{-1-im^2}\exp\big[\quart\tau_P(r)\big]
\end{align}
satisfy the BSE \eq{num10} and \eq{num11} at \order{r^{-2}} for $r \to \infty$ with no condition on $N_8/N_9$. The agreement improves to \order{r^{-3}} at $t=1\ (\xt=0)$.

%%%%%%%%%%%%%%%%%%%%%%%%%%
\subsection{Asymptotic behavior of the $J^{PC} = 0^{-+}$ wave function for $\xt \to \infty$ in region $B$} \label{sec5D2}
%%%%%%%%%%%%%%%%%%%%%%%%%%

Taking $r \to \infty$ at fixed $P$ means $\tau_P(r) = M^2-2EV+V^2 \simeq V^2 >0$, \ie, the limit is defined in region $C$ \eq{e5.1a}. To enable large $r$ in region $B$, where $E-P \leq V'r \leq E+P$, I first take $P \to \infty$ at fixed $\xt$ with $z \propto 1/P$ (Lorentz contraction). Thus $r \to \xt$ and $\tau_P \simeq -2EV'\xt < 0$. The BSE \eq{num10} and \eq{num11} become
\begin{align} 
\phip_4 \simeq \phip_{11}  &\simeq -\frac{1}{V}\big(\dpp\phip_9+m\phip_8\big)  \hspace{3cm}
\phip_{13} \simeq \phip_{14} \simeq -\frac{1}{V}\big(m\phip_9+\inv{\xt}\dpp(\xt\phip_8)\big) \nn\crt
\phip_{10} &\simeq \frac{2}{E-V}\,\dz\phip_9 \hspace{4.3cm}
\phip_{15} \simeq -\frac{2}{E-V}\,\dz\phip_8   \label{e5.5} \crt
\halft(E-V)\phip_{8}-\dz\phip_{15} &= m\phip_4-\dpp\phip_{14} \hspace{2.3cm}
\halft(E-V)\phip_{9}+\dz\phip_{10} = m\phip_{13}- \inv{\xt}\dpp(\xt\phip_{11}) \label{e5.6}
\end{align}
All wave functions are of \order{P^0} since $\dz \propto P$. It is convenient to introduce the uncontracted, \order{P^0} coordinate $z_U$,
\begin{align} \label{e5.7} 
z_U &\equiv \frac{E-V}{M}z  \hspace{2.9cm}  \frac{\partial{z_U}}{\partial z} \simeq \frac{E-V}{M} \nn\crt
\phip_{10} &\simeq \frac{2}{M}\partial_{z_U}\phip_9 \hspace{2.3cm} 
\dz\phip_{10} \simeq \frac{2(E-V)}{M^2}\partial_{z_U}^2\phip_9 \nn\crt
\phip_{15} &\simeq -\frac{2}{M}\partial_{z_U}\phip_8 \hspace{2cm} 
\dz\phip_{15} \simeq -\frac{2(E-V)}{M^2}\partial_{z_U}^2\phip_8
\end{align}
On the lhs. of \eq{e5.6} the \order{P} term must cancel. This motivates to consider
\begin{align} \label{e5.8}  
\phip_k(z,\xt) = \cos(\halft Mz_U)f_k(\xt) \hspace{1cm} (k=8,9)
\end{align}
for which $\phip_k+(4/M^2)\partial_{z_U}^2\phip_k =0$, requiring the rhs. of Eqs. \eq{e5.6} to vanish. At leading order for $\xt\to\infty$ the $\dpp$ derivatives operate only on $f_k(\xt)$, not on the powers of $\xt$ multiplying it. Consequently,
\begin{align} \label{e5.9} 
-V (m\phip_4-\dpp\phip_{14}) &\simeq m\dpp\phip_9+m^2\phip_8-m\dpp\phip_9-\dpp^2\phip_8 
= m^2\phip_8-\dpp^2\phip_8 =0 \nn\crt
-V \big[m\phip_{13}- \inv{\xt}\dpp(\xt\phip_{11})\big] &\simeq m^2\phip_9+m\dpp\phip_8-\dpp^2\phip_9-m\dpp\phip_8 = m^2\phip_9-\dpp^2\phip_9 =0
\end{align}
It follows that the wave functions behave exponentially for $\xt \to \infty$,
\begin{align} \label{e5.10} 
f_k(\xt) \simeq \exp(\pm m\xt)\hspace{1cm} (k=8,9) \hspace{1cm} \mbox{for}\ \xt\to\infty
\end{align}
The physical solution is exponentially suppressed, suggestive of a tunneling behavior.

%%%%%%%%%%%%%%
\subsection{Chebyshev polynomials \label{appB1}}
%%%%%%%%%%%%%%

The $n$:th order Chebyshev polynomials $T_n(x)$ are for $-1\leq x \leq 1$ defined by
\begin{align} \label{num1} 
T_n(\cos\theta) = \cos(n\theta)
\end{align}
Consequently $T_0(x)=1,\ T_1(x)=x,\ T_2(x)=2x^2-1,\ T_3(x)=4x^3-3x$, and
\begin{align} \label{num2} 
T_n&(-x) = (-1)^n T_n(x)  \nn\crt
T_n(1)=1& \hspace{2cm} T_n(-1)=(-1)^n \nn\crt
T_{2n}(0) = (-1)^n& \hspace{2cm} T_{2n+1}(0) = 0 \nn\crt
\partial_x T_n(x)\Big|_{x=1} = n^2& \hspace{2cm} \partial_x T_n(x)\Big|_{x=-1}= n^2(-1)^{n+1} \nn\crt
\partial_x^2 T_n(x)\Big|_{x=1} = \inv{3}n^2(n^2-1)& \hspace{2cm} \partial_x^2 T_n(x)\Big|_{x=-1}= \inv{3}n^2(n^2-1)(-1)^{n}
\end{align}

\bibliography{Boost_23_refs.bib}
\end{document}